\documentclass[12pt]{article}
\usepackage{amsmath,amsfonts,graphicx,color,bbm,tikz,float,mathrsfs,amssymb,xcolor}
\usepackage{comment}
\usepackage[nosort]{cite}
\usepackage{subfigure}
\usetikzlibrary{calc,positioning}
\usetikzlibrary{patterns,arrows,decorations.pathreplacing}
\usepackage{caption}
\usepackage{ulem}
\tikzset{>=stealth}
\usepackage{lipsum}
\usepackage{tikz}

\makeatletter\@addtoreset{equation}{section}\makeatother
\newcommand{\be}{\begin{equation}}
\newcommand{\ee}{\end{equation}}
\def\beq{\begin{equation}}
\def\eeq{\end{equation}}
\newcommand{\bea}{\begin{eqnarray}}
\newcommand{\eea}{\end{eqnarray}}
\newcommand{\Tr}{{\rm Tr\,}}

\newcommand{\cA}{{\cal A}}

\newcommand{\cN}{{\cal N}}
\newcommand{\cP}{{\cal P}}

\newcommand{\cH}{{\cal H}}

\newcommand{\bra}[1]{{\left< {#1} \right|}}
\newcommand{\ket}[1]{{\left| {#1} \right>}}

\def\nn{\nonumber}

\renewcommand{\title}[1]{\vbox{\center\LARGE{#1}}\vspace{3mm}}
\renewcommand{\author}[1]{\vbox{\center#1}\vspace{3mm}}

\newcommand{\email}[1]{\vbox{\center\tt#1}\vspace{3mm}}

\newcommand{\jti}{\tilde{j}}
\newcommand{\Lti}{\tilde{L}}
\newcommand{\pti}{\tilde{p}}
\newcommand{\kti}{\tilde{k}}
\newcommand{\bX}{{\bf X}}

\begin{document}
\begin{titlepage}
\begin{center}

{\large {\bf 
Novel quantum phases on graphs using abelian gauge theory} }

\author{Pramod Padmanabhan,$^a$ Fumihiko Sugino$^b$ }

\vskip 0.1cm
{$^a${\it School of Basic Sciences,\\ 
Indian Institute of Technology, Bhubaneswar, India}}
\vskip0.1cm
{ $^b${\it Center for Theoretical Physics of the Universe,\\
Institute for Basic Science, Daejeon, South Korea}} 

\email{pramod23phys, fusugino@gmail.com}

\vskip 1cm 
\end{center}

\abstract{
\noindent 
Graphs are topological spaces that include broader objects than discretized manifolds, making them interesting playgrounds for the study of quantum phases not realized by symmetry breaking. 
In particular they are known to support anyons of an even richer variety than the two-dimensional space. 
We explore this possibility by building a class of frustration-free and gapped Hamiltonians based on discrete abelian gauge groups. 
The resulting models have a ground state degeneracy that can be either a topological invariant,
an extensive quantity or a mixture of the two. 
For two basis of the degenerate ground states which are complementary in quantum theory, the entanglement entropy is exactly computed. 
The result for one basis has a constant global term, known as the topological entanglement entropy, implying long-range entanglement. 
On the other hand, the topological entanglement entropy vanishes in the result for the other basis. Comparisons are made with similar occurrences in the toric code. 
We analyze excitations and identify anyon-like excitations that account for the topological entanglement entropy. 
An analogy between the ground states of this system and the $\theta$-vacuum for a $U(1)$ gauge theory on a circle is also drawn. 
}

\end{titlepage}
\tableofcontents 

\section{Introduction}
\label{Introduction}

Quantum phases of matter, or the ones that are beyond Landau's classification of spontaneous symmetry breaking with local order parameters, 
have gained theoretical and experimental significance over the last few decades. 
Among them, the so called {\it topological phases of matter} or the {\it topologically ordered} phases have become increasingly important for robust methods of quantum computation \cite{Nayak, Freedman, Pachos}. While topological order can manifest itself in different ways in different dimensions, 
a class of solvable examples in two dimensions are given by the quantum double models of Kitaev \cite{kitaev}.  
They are characterized by a ground state degeneracy (GSD) that is a topological invariant, anyonic excitations, and ground state entanglement entropy (EE) that includes a global component depending on the superselection sectors of the theory \cite{kp} as a subleading order term. The quantum double models are Hamiltonian realizations of discrete gauge theories or gauge theories based on finite groups which were well studied in the early 90's \cite{bais}. More generally they can also be considered for {\it involutory Hopf algebras} \cite{majid}, of which the group algebra is a special case \cite{kitaev, bomb, pp2}.

Most studies consider long-ranged topologically ordered systems in two and three dimensions. 
The models predominantly are located on lattices that discretize a two- or three-dimensional differentiable manifold. 
In this paper we construct exactly solvable models for quantum phases on {\it connected graphs} which do not fall into the usual setups for physical systems, since graphs include broader objects than discretized manifolds. 
However a graph is still a {\it topological space} and can be conveniently thought of as a one-dimensional CW complex \cite{hatcher}. In fact physics on graphs or networks, as it is sometimes called in the literature, can be rather non-trivial, with early works studying the issue of particle statistics on such spaces \cite{bal}.\footnote{Graph structures also appear in the physics and math literature under the name of {\it quantum graphs} especially in the area of mesoscopic physics \cite{kuch1, kuch2, ingr}.} 
More importantly, there have been studies exploring the possibilities of anyons on graphs, both abelian and non-abelian ones~\cite{alicea, agraph1, agraph11, agraph12}. 
Analogous to {\it braid groups} being the fundamental groups of the configuration space of $N$ identical particles in $\mathbb{R}^2$ \cite{lm}, {\it graph braid groups} play a similar role on 
different types of graphs \cite{agraph2, agraph3, agraph4}. 
However, these have a fundamental difference from the conventional braid groups, as the generators do not obey a Yang-Baxter type relation.

It is reasonable to expect that our models on graphs share some features with systems of the known topologically ordered phases.   
The goal of this paper is to explore this possibility by analyzing models based on discrete abelian gauge groups, 
which are quite similar in form to the abelian quantum double models or the toric code. 
The main difference is that now we also include `matter' fields on the vertices of the lattice or graph in addition to the gauge fields on the edges/links of the lattice or graph. 
While the vertex operators or the gauge transformations of the toric code are slightly modified to act on the matter fields on the vertices as well, 
the plaquette operators or the operators measuring local flux of the toric code are replaced by an entirely new operator known as the {\it edge operator}. 
Before we go into the details of the models we would like to emphasize that the models presented here can be obtained from \cite{pt1, pt2, pp1, mar1, mar2} where topological order is discussed from the point of view of {\it higher gauge theories} \cite{baez} constructed using 2-groups and other higher categories. 
The papers \cite{pt1, pt2} study topological order in various dimensions using simplicial complexes and in this context what we present here is a detailed study of the models in the simplest such complex, namely a graph. 

The rest of this paper is organized as follows. 
The operators on graphs, including the Hamiltonian, are defined in section~\ref{Operators}. 
The models are parametrized by two integers, $m$ and $n$, which are the dimensions of the local Hilbert spaces on the vertices and edges of the graph, respectively. 
Following this we cover all the tell-tale signs of topological order starting with a detailed analysis of the ground states in section~\ref{sec:GS}. 
For general $m$ and $n$, we find the GSD to be a function of a topological invariant (the first Betti number) and a graph invariant (the number of vertices). 
The latter gives an extensive dependence on the system size.\footnote{
Graph invariants are invariant quantities under graph isomorphisms. 
There are more non-trivial graph invariants known as {\it Tutte polynomials} which also arise in statistical physics \cite{tutte}.}
Next we exactly compute the EE of these ground states in section~\ref{sec:EE}, and find that there is a global constant term known as the topological EE, 
which exists regardless of the partition of the graph. 
We also compute the EE for superpositions of the ground states that are complementary to the previous ground states, in which the topological EE turns out to vanish. 
Different aspects from arguments on the {\it minimal entropy states} given in~\cite{asvin,wang1} are observed here.
In section~\ref{sec:Estates} we see that the total quantum dimension of the system obtained from the topological EE is precisely equal to the number of anyon-like excitations.  
In section~\ref{sec:Disc} we summarize the result and discuss some future directions. 
In appendix~\ref{app:analogy}, we present an analogy of the model to quantum field theory with $U(1)$ gauge field and matter field to gain an intuitive understanding. 
In appendix~\ref{app:U(1)}, $U(1)$ gauge theory on a circle obeying twisted boundary conditions is briefly discussed 
to help understand the ground states of our models. Appendix~\ref{app:betti} is devoted to some topological aspects of graphs.   

\section{The operators}
\label{Operators}

Let $G=(V,E)$ be a connected graph composed by a set of vertices $V$ and a set of edges $E$. Each edge is endowed with an orientation and its endpoints are attached to vertices in $V$. For each of such graphs, the adjacency matrix is well-defined.\footnote{
The $(i,j)$-th matrix element stands for the number of edges directed from the $i$-th vertex to the $j$-th vertex. 
The diagonal $(i,i)$-th matrix element counts the number of self-loops at the $i$-th vertex.}  

We place finite dimensional Hilbert spaces on both the vertices and the edges making the total Hilbert space, $\mathcal{H}=\otimes_{v\in V}\mathcal{H}_v\otimes_{e\in E}\mathcal{H}_e$. 
Upon taking 
$\mathcal{H}_v = \textrm{Span of}\,\{\ket{h_v} | h_v=0,1,\cdots, m-1\}\simeq\mathbb{C}^m$
for each $v$ and 
$\mathcal{H}_e= \textrm{Span of}\,\{\ket{i_e} | i_e=0,1,\cdots, n-1\}\simeq\mathbb{C}^n$
for each $e$, we can consider the local Hilbert spaces as carrying the representations of the abelian groups, $\mathbb{Z}_m$ and $\mathbb{Z}_n$ respectively. In the parlance of many-body physics these are $m$ and $n$ level systems or spins $\frac{m-1}{2}$ and $\frac{n-1}{2}$ on the vertices and edges respectively. 

Furthermore we consider a homomorphism, $\partial : \mathbb{Z}_n\rightarrow\mathbb{Z}_m$. 
For  any $\mathbb{Z}_n$ elements $a,\,b\in\{0,1,\cdots, n-1\}$, the homomorphism satisfies $\partial(a+b)=\partial(a) + \partial(b)$. 
The $\mathbb{Z}_n$ degrees of freedom on each edge are regarded as `gauge fields', and the $\mathbb{Z}_m$ degrees of freedoms on each vertex as `matter fields'. 
The homomorphism induces the gauge transformation property of the matter fields from that of the gauge fields.    
Let $k$ be the greatest common divisor of $m$ and $n$ ($\gcd(m,n)=k$). Then, we can write $m$ and $n$ as 
\be
m=kp,\qquad n=kq,
\label{mnpq}
\ee
where $p$ and $q$ are coprime integers ($\gcd(p,q)=1$). 
The possible choices for the homomorphism are labelled by the group $\mathbb{Z}_k$ 
and given by 
\begin{equation}\label{homo}
\partial^{\left[l\right]}(j) = pjl 
\end{equation}
with $l\in \mathbb{Z}_k=\{0, 1, \cdots,k-1\}$ labelling the homomorphisms. 
These give compatible homomorphisms as $\partial^{\left[l\right]}(n) = pnl = m$, under mod $m$ arithmetic, 
implying that $\partial^{\left[l\right]}$'s map the identity of $\mathbb{Z}_n$ to the identity of $\mathbb{Z}_m$.
For later convenience, we also introduce the greatest common divisor of $k$ and $p$ which is denoted by $\xi$: $\gcd(k,p)=\xi$. Namely, 
\be
k=\xi\kti, \qquad p=\xi\pti \qquad \textrm{with} \qquad \gcd(\kti,\pti)=1.
\label{gcd_kti_pti}
\ee 

Using these ingredients we define the operators that make up the Hamiltonian. 
Although it is possible for a general $l$, the case $l=1$ is mainly considered in what follows for simplicity. Then (\ref{homo}) becomes $\partial(j)=pj$. 

\subsection{Vertex operator or gauge transformations} 
\label{sec:vo}
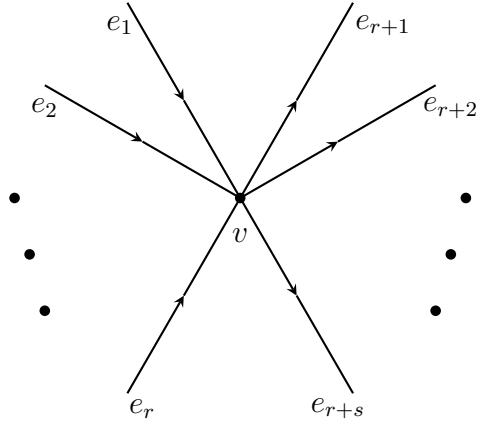
\begin{figure}[H]
\centering
\captionsetup{width=.8\linewidth}
\begin{tikzpicture}
%
\fill (0,0) circle [radius=2pt];
\draw [->, thick] (0,0)--(1.299, 0.75);
\draw [-, thick] (1.299, 0.75)--(2.598, 1.5);
\draw [->, thick] (0,0)--(0.75, 1.299);
\draw [-, thick] (0.75, 1.299)--(1.5, 2.598);
\fill (3,0) circle [radius=2pt];
\fill (2.8,-0.75) circle [radius=2pt];
\fill (2.598, -1.5) circle [radius=2pt];
\draw [->, thick] (0,0)--(0.75, -1.299);
\draw [-, thick] (0.75, -1.299)--(1.5, -2.598);
\draw [-, thick] (0,0)--(-1.299, 0.75);
\draw [->, thick] (-2.598, 1.5)--(-1.299, 0.75);
\draw [-, thick] (0,0)--(-0.75, 1.299);
\draw [->, thick] (-1.5, 2.598)--(-0.75, 1.299);
\fill (-3,0) circle [radius=2pt];
\fill (-2.8,-0.75) circle [radius=2pt];
\fill (-2.598, -1.5) circle [radius=2pt];
\draw [-, thick ] (0,0)--(-0.75, -1.299);
\draw [->, thick] (-1.5, -2.598)--(-0.75, -1.299);
\node (v) at (0,-0.5) {$v$};
\node (e1) at (-1.6, 2.298) {$e_1$};
\node (e2) at (-2.598,1.2) {$e_2$};
\node (er) at (-1.3,-2.798) {$e_r$};
\node (er1) at (1.9, 2.298) {$e_{r+1}$};
\node (er2) at (2.798,1.2) {$e_{r+2}$};
\node (ers) at (1.3,-2.798) {$e_{r+s}$};
\end{tikzpicture}
\caption{A vertex $v$ and attached $r+s$ edges. The left $r$ edges ($e_1, e_2, \cdots, e_r$) are directed to the vertex, 
whereas the right $s$ edges ($e_{r+1}, e_{r+2}, \cdots, e_{r+s}$) are outgoing from the vertex. 
}
\label{vertex}
\end{figure}

The {\it vertex operator}, $A_v$, implements the gauge transformations 
of the gauge group $\mathbb{Z}_n$, and acts nontrivially on 
$\cH_v$ and $\cH_e$ with the edges $e$ attached to $v$. 
Let $L_v$ be the set of edges attached to the vertex $v$. $L_v$ is divided into a set of edges directed inwards to $v$, $L_v^+$, and a set of edges directed outwards from $v$, $L_v^-$: $L_v=L_v^+ \cup L_v^-$.
For the example depicted in Fig.~\ref{vertex}, $L_v^+=\{e_1,\cdots, e_r\}$ and $L_v^-=\{e_{r+1},\cdots, e_{r+s}\}$. 
The operator $A_v$ is defined as 
\begin{equation}\label{vo}
A_v = \frac{1}{n}\sum\limits_{j=0}^{n-1} A_v^{(j)} ,
\qquad
A_v^{(j)} \equiv  x_v^{\partial(j)}\,\bX_{L_v}^{pj}
\ee
with
\be
\bX_{L_v}\equiv \left(\prod_{e\in L_v^+}X_{e}\right)\,\left(\prod_{e\in L_v^-}X_{e}^{-1}\right).
\ee
$x_v$ and $X_e$ are the shift operators on the basis of $\mathbb{C}^m$ and $\mathbb{C}^n$ respectively:
\be
x_v\ket{h_v}=\ket{h_v+1}, \qquad
X_e\ket{i_e} = \ket{i_e+1}, 
\label{shift}
\ee
where the numbers $h_v$ and $i_e$ are evaluated in mod $m$ and mod $n$ arithmetic respectively. 
For the example in Fig.~\ref{vertex}, $A_v^{(j)}=x_v^{\partial(j)}\,\left(\prod_{a=1}^rX_{e_a}^{pj}\right)\,\left(\prod_{b=r+1}^{r+s}X_{e_b}^{-pj}\right)$ acts on the local Hilbert spaces as
\begin{equation}\label{vjl}
A_v^{(j)} \left(\ket{h_v}\prod_{a=1}^r\ket{i_{e_a}}\prod_{b=r+1}^{r+s} \ket{i_{e_b}}\right)
= \ket{h_v+pj}\prod_{a=1}^r\ket{i_{e_a}+pj}\prod_{b=r+1}^{r+s}\ket{i_{e_b}-pj}.
\end{equation} 

The vertex operator (\ref{vo}) is easily seen to be a projector $\left(A_v\right)^2=A_v$, as it is a {\it  group average} over $\mathbb{Z}_n$. 
It has $n-1$ other mutually orthogonal projectors\footnote{These properties follow from the {\it orthogonality theorem} for group characters, also known as the {\it Schur orthogonality relations} \cite{hammer}.} that are labelled by the {\it irreducible} representations (IRRs) of $\mathbb{Z}_n$,
\begin{equation}\label{voperps}
A_v^{\left[\alpha\right]} = \frac{1}{n}\sum\limits_{j=0}^{n-1} \chi_{\alpha,n}(j)A_v^{(j)},
\end{equation}
where $\alpha$ labels the IRR and $\chi_{\alpha,n}(j)$ is the character of the element $j\in\mathbb{Z}_n$ in the IRR $\alpha$. 
Explicitly, $\chi_{\alpha,n}(j) = \omega_n^{\alpha j}$ with $\alpha\in\{0,1,\cdots,n-1\}$. 
Here and in what follows, $\omega_d\equiv e^{\frac{2\pi\mathrm{i}}{d}}$ for a positive integer $d$. 
In this notation the vertex operator in (\ref{vo}) corresponds to the trivial IRR ($\alpha=0$). 

\subsection{Edge operators or 0-holonomy operators} 
The plaquette operators in the toric code or more generally the quantum double models \cite{kitaev} measure the flux of the gauge fields around a plaquette 
(or in other words  the smallest  {\it Wilson loop}) for the discrete gauge group. 
We call this the {\it 1-holonomy} operator.\footnote{We use 0-holonomy and 1-holonomy keeping in mind that these operators can be generalized to abelian {\it higher gauge groups} as in \cite{pt1}. In the language of higher gauge theory as described in \cite{pt1}, matter fields on vertices are 0-gauge fields and the gauge fields on edges are 1-gauge fields. This can be generalized to $d$-gauge fields living on $d$-dimensional simplices of a simplicial complex.} 
In a similar manner, we consider the `0-holonomy' operator or edge operator which acts on two adjacent vertices and the link in-between as in Fig.~\ref{edge}. 
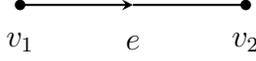
\begin{figure}[H]
\centering
\captionsetup{width=.8\linewidth}
\begin{tikzpicture}
\draw[->,thick] (0,0)--(1.5,0);
\draw[-,thick] (1.5,0)--(2.95,0);
\fill (0,0) circle [radius=2pt];
\fill (3,0) circle [radius=2pt];
\node (v1) at (0,-0.5) {$v_1$};
\node (v2) at (3,-0.5) {$v_2$};
\node (e) at (1.5,-0.5) {$e$};   
\end{tikzpicture}
\caption{A directed edge $e$ flanked by vertices $v_1$ and $v_2$.}
	\label{edge}
\end{figure}
\noindent
It is defined as\footnote{
 Due to the presence of the gauge field on $e$ this operator is sometimes called as the `fake' 0-holonomy in the literature \cite{pt1, pt2, mar1, mar2}. Also, when the vertices $v_1$ and $v_2$ coincide ($v_1=v_2\equiv v$) 
 and the edge $e$ forms a self-loop, $B_e^{(j)}$ becomes 
 $B_e^{(j)}= 1_v Z_e^{qj}$ with $1_v$ the identity operator on $\cH_v$. 
 }
\be
B_e=\frac{1}{k}\sum_{j=0}^{k-1}B_e^{(j)}
\qquad
\mbox{with}
\qquad
B_e^{(j)} \equiv z_{v_1}^{pj}\,Z_e^{qj}\,z_{v_2}^{-pj},
\label{eo}
\ee
where $z_v$ and $Z_e$ are clock operators on the basis of $\mathbb{C}^m$ and $\mathbb{C}^n$ respectively:
\begin{equation}
z_v\ket{h_v} = \omega_m^{h_v}\ket{h_v},\qquad Z_e\ket{i_e} = \omega_n^{i_e}\ket{i_e}.
\label{clock}
\end{equation}
It is easy to see that $B_e$ is a projector $\left(B_e\right)^2=B_e$ and is diagonal on the basis. 
It turns out that the edge operator or 0-holonomy operator (\ref{eo}) acts on the local Hilbert spaces as 
\begin{equation}\label{0hol}
B_e \ket{h_{v_1}}\ket{h_{v_2}}\ket{i_e} = \delta_m\left(p\left(h_{v_2}-h_{v_1}\right), \partial(i_e)\right) \ket{h_{v_1}}\ket{h_{v_2}}\ket{i_e} 
\end{equation}
with $\delta_m(a,b)$ being the mod $m$ Kronecker delta for integers $a$ and $b$. 
Physically this can be regarded as measuring the 0-flux due to the matter fields or the 0-gauge fields across the 1-gauge field on the edge~$e$. 

As with the vertex operators, we can write down the orthogonal projectors of these edge operators by projecting to different IRRs of $\mathbb{Z}_m$ as
\begin{equation}\label{orthoBe}
B_e^{\left[\alpha\right]} = \frac{1}{k}\sum\limits_{j=0}^{k-1} \chi_{\alpha,m}(pj)\,B_e^{(j)}. 
\end{equation}
Since $\chi_{\alpha,m}(pj)\equiv \omega_m^{\alpha pj}=\omega_k^{\alpha j}$, independent operators are given by $\alpha\in\mathbb{Z}_k=\{0,1,\cdots, k-1\}$ rather than $\mathbb{Z}_m$. 

\subsection{The Hamiltonian} 
\label{sec:H}
For a graph $G=(E,V)$, the Hamiltonian $H$ is constructed out of the vertex and edge operators (\ref{vo}) and (\ref{eo}) as\footnote{
For a general choice of the homomorphism $l$, the vertex operator (\ref{vo}) with $\partial(j)$ replaced by $\partial^{[l]}(j)$ and the edge operator (\ref{eo}) with $Z_e^{qj}$ changed by 
$Z_e^{qjl}$, compose the Hamiltonian. \label{ft:homol}
}
\begin{equation}\label{H}
H = -\sum\limits_{v\in V} A_v - \sum\limits_{e\in E} B_e.
\end{equation}
From the properties of the shift and clock operators
\be
x_v^{j_1}z_v^{j_2}=\omega_m^{-j_1j_2}z_v^{j_2}x_v^{j_1} \qquad \mbox{and} \qquad 
X_e^{j_1}Z_e^{j_2}=\omega_n^{-j_1j_2}Z_e^{j_2}X_e^{j_1},
\label{xz_XZ}
\ee
it is easy to show that projectors $A_v$ and $B_e$ mutually commute with each other:
\be
\left[A_v,\,B_e\right]= \left[A_v,\,A_{v'}\right]= \left[B_e,\,B_{e'}\right]=0
\ee
for any $v,\,v'\in V$ and $e,\,e'\in E$.\footnote{This can also be seen from the actions on the local Hilbert spaces (\ref{voperps}) and (\ref{0hol}), which can be used for generalizations to arbitrary finite non-abelian groups as well \cite{pp1}.
} 
The operators $A_v$ and $B_e$ either have no overlap or share one edge and one vertex. In the latter case, noncommutativity of the operators on the edge cancels with that of the operators at the vertex, 
which is analogous to the abelian toric code models \cite{kitaev}.
Thus the Hamiltonian (\ref{H}) is a sum of commuting projectors, and hence it is gapped and frustration-free.

Note that the operators in the Hamiltonian are well-defined for an arbitrary graph irrespective of it being planar or non-planar. 
Although we only consider the case where the gauge group is $\mathbb{Z}_n$ and the matter fields belong to $\mathbb{Z}_m$, the model can be extended to an arbitrary abelian group. 
An analogy to quantum field theory with $U(1)$ gauge field and a matter field is presented in appendix~\ref{app:analogy}. 

We mention some symmetry properties of the system:
\begin{itemize} 
\item{\bf Local symmetries} - It is easy to verify that $x_v\bX_{L_v}$ and $X^k_e$ for any $v\in V$ and $e\in E$ commute with the Hamiltonian (\ref{H}), 
and generate local $\mathbb{Z}_{kpq}$ and $\mathbb{Z}_q$ symmetries respectively. 
$z_v^k$ and $Z_e^{\tilde{n}}$ with
\be
\tilde{n}\equiv \frac{n}{\xi}=\kti q
\label{ntilde}
\ee
are also local operators commuting with the Hamiltonian ($\xi$ is the greatest common 
divisor of $k$ and $p$ as in (\ref{gcd_kti_pti})).

Seemingly $x_v^{k}$ for any $v\in V$ is an additional local $\mathbb{Z}_p$ symmetry transformation, but it is equivalent to one of the above transformations: 
$\left(x_v\bX_{L_v}\right)^{kq}=x_v^{kq}$ due to $\gcd(p,q)=1$.

\item{\bf Quasi-local symmetries} - Consider traversing a closed path $C$, consisting of edges in $E$, in either the clockwise or counterclockwise direction. Then we can write down the operator 
\be
Z(C)\equiv\prod\limits_{e\in C}Z_{e}^{(e|C)}, 
\label{ZC}
\ee
where 
$(e|C)$ is a sign factor according the orientations:  
$(e|C)=1$ for $e$ and $C$ parallel, and $(e|C)=-1$ for $e$ and $C$ anti-parallel. 
This operator is analogous to the Wilson loop of the gauge theory, and commutes with the Hamiltonian (\ref{H}). The number of such independent operators is equal to the number of independent closed paths on the graph: $|E|-|V|+1$, where $|E|$ and $|V|$ are the numbers of edges and vertices in the graph. 
This is equal to the {\it first Betti number}, a topological invariant of the graph.\footnote{Topological aspects of the first Betti number are provided with a brief look at graph homology theory 
in appendix~\ref{app:betti}.} For planar graphs, the first Betti number can be interpreted as the number of one-dimensional holes. 

\item{\bf Global symmetries} - The operator, $\prod\limits_{v\in V}x_v$, with support spanning all the vertices of the graph commutes with the Hamiltonian (\ref{H}) 
and generates a global $\mathbb{Z}_m$ symmetry. 
This is deduced from the above local symmetry as $\prod_{v\in V}x_v\bX_{L_v}$ upon using the constraint $\prod_{v\in V} \bX_{L_v}=1$.  
The Hamiltonian is also invariant under parity, which is realized on a directed graph by reversing the orientations of all the edges, seen via taking the inverse of the shift and clock operators on all vertices and edges.
\end{itemize}

Notice that 
the local transformation $X^k_e$ does not commute with the quasi-local transformation (\ref{ZC}) when $e$ is on the path $C$. Thus, any state cannot respect both of the two symmetries. 
Similarly, $x_v\bX_{L_v}$ does not commute with $z_v^k$ and $Z_e^{\tilde{n}}$. 


\section{Ground states}
\label{sec:GS}
The Hamiltonian (\ref{H}) is exactly solvable. Since the $\textrm{Spec}(A_v) = \textrm{Spec}(B_e) = \{0,1\}$ for all $v$ and $e$, the lowest energy is given by $E_0=-|V|-|E|$, and any ground state $\ket{\mbox{GS}}$ 
satisfies $A_v\ket{\mbox{GS}}=B_e\ket{\mbox{GS}}=\ket{\mbox{GS}}$ for all $v\in V$ and $e\in E$. From this it follows that the projector to the ground state manifold is given by 
\begin{equation}
\pi_0 = \prod\limits_{v\in V}A_v\prod\limits_{e\in E}B_e.
\label{pi0}
\end{equation}
It is clear then that the GSD is given by the trace of $\pi_0$ over the total Hilbert space:
\begin{equation}\label{gsd}
\textrm{GSD} = \Tr_{\cH}\left(\pi_0\right).
\end{equation}

\subsection{GSD}
\label{sec:GSD}
To compute the trace we observe that the non-zero contribution comes from the term $\prod\limits_{v\in V}1_v\prod\limits_{e\in E}1_e$ in $\pi_0$, 
where $1_v$ and $1_e$ are the identity operators on $\mathcal{H}_v$ and $\mathcal{H}_e$ respectively. 
Note that $\Tr_{\cH}=\left(\prod_{v\in V}\Tr_v\right)\left(\prod_{e\in E}\Tr_e\right)$ with $\Tr_v$ and $\Tr_e$ being the traces over the Hilbert spaces $\cH_v$ and $\cH_e$. 
For the shift and clock operators, $x_v^m=z_v^m=1_v$ and $x_v^az_v^b$ is traceless unless $a,b\in m\mathbb{Z}$. 
The same holds for $X_e$ and $Z_e$. 
Then, from (\ref{vo}), (\ref{eo}) and (\ref{pi0}), $\pi_0$ is written as a polynomial of $x_v$, $z_v$, $X_e$ and $Z_e$ for all $v\in V$, $e\in E$. 
Among the terms of the polynomial, only the term proportional to the identity $\prod\limits_{v\in V}1_v\prod\limits_{e\in E}1_e$ gives non-zero contribution under the trace operation in (\ref{gsd}). 
Any other term contains 
at least one of $x_v$, $z_v$, $X_e$ and $Z_e$ for some $v\in V$, $e\in E$, and vanishes under the trace. 

Thus computing the GSD translates into an exercise of counting the number of $\prod\limits_{v\in V}1_v\prod\limits_{e\in E}1_e$'s in $\pi_0$. 
The answer should be the multiplication of the three factors: 
\begin{itemize}
\item
$\left(\frac{1}{n}\right)^{|V|}\left(\frac{1}{k}\right)^{|E|}$ from the prefactors of the sums in $A_v$ (\ref{vo}) and $B_e$ (\ref{eo})
\item
the dimension of the total Hilbert space $m^{|V|}n^{|E|}$
\item
the number of $\prod\limits_{v\in V}1_v\prod\limits_{e\in E}1_e$'s. 
\end{itemize}
 
When $j=k,\,2k, \cdots, (q-1)k$ the operator $x_v^{pj}$ in $A_v^{(j)}$ (\ref{vo}) becomes the identity, but the operators on the edges in $A_v^{(j)}$ remain nontrivial. 
Note that $pj/k\neq 0$ (mod $q$) because of $\gcd(p,q)=1$.  
Also, for $j=1,\,2,\cdots, k-1$, $Z_e^{qj}$ in $B_e^{(j)}$ (\ref{eo}) is always nontrivial, which physically implies that there is no matter field with neutral electric charge. 
These properties lead to\footnote{
The result (\ref{GSDresult}) remains valid for the homomorphism with other choice of $l$, as long as $Z_e^{qjl}$ in the edge operator (see footnote~\ref{ft:homol}) 
is not trivial for any $j\in\{1,\,2,\,\cdots,k-1\}$. 
Otherwise, matter fields with neutral electric charge appear, and the GSD would depend on other details of the graph in addition to $|V|$ and $|E|$.}  
  \bea
\mbox{GSD} & = & \Tr_{\cH}\left(\prod_{v\in V}A_v\prod_{e\in E}B_e\right) \nn \\
& = & \left(\frac{1}{n}\right)^{|V|}\left(\frac{1}{k}\right)^{|E|}\times m^{|V|}n^{|E|}\times q \nn \\
& = & p^{|V|}q^{B_1},
\label{GSDresult}
\eea
where $B_1=|E|-|V|+1$ is the first Betti number of the graph. For any fixed $j\in k\mathbb{Z}_q=\{0,\,k,\,2k,\,\cdots, (q-1)k\}$, the relevant contribution to GSD solely comes from $A_v^{(j)}$ and $B_e^{(0)}$ for all 
$v\in V$ and $e\in E$, making up $\prod\limits_{v\in V}1_v\prod\limits_{e\in E}1_e$. 
Note that for an edge $e$ connecting two vertices $v_1$ and $v_2$ as in Fig~\ref{edge}, $A_{v_1}^{(j)}A_{v_2}^{(j)}$ acts trivially on $\cH_e$.  
The last factor $q$ on the second line counts the possible choice of $j$ that is the number of $\prod\limits_{v\in V}1_v\prod\limits_{e\in E}1_e$'s.

\subsection{Construction of ground states}
\label{sec:GSs}
If we find a seed state $\ket{s}$ satisfying $B_e\ket{s}=\ket{s}$ for any $e$,  
one of the ground states is given by 
\be
\ket{\mbox{GS}\,s} = \sqrt{\cN}\,\left(\prod_{v\in V} A_v\right)\ket{s}, 
\label{GSs}
\ee
where $\cN$ is a normalization constant. 
It is easy to see that $\ket{s=0}\equiv\prod_{v\in V}\ket{0_v}\prod_{e\in E}\ket{0_e}$ gives such a state $\ket{s}$. 

Starting with a ground state
\be
\ket{\mbox{GS}\,0}\equiv \sqrt{\cN}\,\left(\prod_{v\in V} A_v\right)\ket{s=0},
\ee
we can exhaust the other ground states by acting the local operators $x_v^a{\bf X}_{L_v}^a$ ($a\in \mathbb{Z}_p$) and $X_{e}^{bk}$ ($b\in \mathbb{Z}_q$) on $\ket{s=0}$. 
Note that $x_v\bX_{L_v}$ acts as a $\mathbb{Z}_p$-transformation on $\ket{\mbox{GS}\,0}$ or on $A_v$, because of $x_v^p\bX_{L_v}^p A_v= A_v$. 
However, all the choices are not independent. 
For 
\be
A_v^{(bk)}=1_v\bX_{L_v}^{pbk} 
\label{Avbk}
\ee
in (\ref{vo}), $A_vA_v^{(bk)}=A_v$ ($b\in \mathbb{Z}_q$) holds, which means that  $\ket{s_1}$ and $\ket{s_2}$ such that $\ket{s_1}=A_v^{(bk)}\ket{s_2}$ for any $b$ and $v$ give the same ground state. 
Taking into account the constraint $\prod_{v\in V}A_v^{(bk)}=1$, the number of the independent choices amounts to 
\be
\frac{p^{|V|}q^{|E|}}{q^{|V|-1}}=p^{|V|}q^{B_1}=\mbox{GSD}.
\ee
Alternatively, we pick edges $\hat{e}_L$ ($L=1,\cdots, B_1$) such that 
the graph $T\equiv G-\{\hat{e}_1,\cdots,\hat{e}_{B_1}\}$ becomes a connected tree graph, i.e., a spanning tree.  
Then, the independent ground states are generated by acting $x_v^{a_v}{\bf X}_{L_v}^{a_v}$ ($a_v\in \mathbb{Z}_p$, $v\in V$) and $X_{\hat{e}_L}^{b_Lk}$ ($b_L\in \mathbb{Z}_q$, $L=1,\cdots,B_1$) on $\ket{s=0}$. 
The states $\ket{s}$ giving the independent ground states are labelled by $\{a_v\left|\right. v\in V\}$ and $\{b_L\left|\right. L=1,\cdots,B_1\}$ for a choice of $\hat{e}_L$'s.   
The choice of $\hat{e}_L$'s is not unique. If one of the $\hat{e}_L$'s, say $\hat{e}_1$, is added to the above spanning tree, a closed path including $\hat{e}_1$ appears. 
We can choose any other edge on the closed path instead of $\hat{e}_1$. 
For example, in a graph depicted in Fig.~\ref{fig:G_ehL},  
we can choose one among $e_{12}$, $e_{37}$, $e_{14}$, $e_{45}$, $e_{56}$ and $e_{67}$ instead of $\hat{e}_1$, 
where $e_{ij}$ denotes the edge in $T$ connecting the vertices $v_i$ and $v_j$. 
The choice of the other $\hat{e}_L$'s can be changed similarly.   
We obtain the same set of independent ground states irrespective of the choice as we see below. 

\begin{figure}[h]
\centering
\captionsetup{width=.8\linewidth}
\begin{tikzpicture}
\draw[->,thick](0,4)--(0,3);
\draw[-,thick] (0,3)--(0,2);
\draw[red,->,thick](0,2)--(0,1);
\draw[red,-,thick] (0,1)--(0,0);
\draw[->,thick](0,0)--(1,-0.5);
\draw[-,thick] (1,-0.5)--(2,-1);
\draw[->,thick](2,5)--(1,4.5);
\draw[-,thick] (1,4.5)--(0,4);
\draw[->,thick](2,5)--(2,4);
\draw[-,thick] (2,4)--(2,3);
\draw[->,thick](2,3)--(2,2);
\draw[-,thick] (2,2)--(2,1);
\draw[->,thick](2,1)--(2,0);
\draw[-,thick] (2,0)--(2,-1);
\draw[->,thick](4,4)--(3,4.5);
\draw[-,thick] (3,4.5)--(2,5);
\draw[red,->,thick](2,3)--(3,3.5);
\draw[red,-,thick] (3,3.5)--(4,4);
\draw[red,->,thick](4,0)--(3,0.5);
\draw[red,-,thick] (3,0.5)--(2,1);
\draw[red,->,thick](2,-1)--(3,-0.5);
\draw[red,-,thick] (3,-0.5)--(4,0);
\draw[red,->,thick](6,-1)--(6,0);
\draw[red,-,thick] (6,0)--(6,1);
\draw[->,thick](6,1)--(6,2);
\draw[-,thick] (6,2)--(6,3);
\draw[->,thick](6,3)--(5,3.5);
\draw[-,thick] (5,3.5)--(4,4);
\draw[red,->,thick] (6,3) arc (0:60:1.3cm);
\draw[red,-,thick] (5.4,4.1) arc (60:123:1.3cm);
\draw[->,thick](6,1)--(5,0.5);
\draw[-,thick] (5,0.5)--(4,0);
\draw[->,thick](4,0)--(5,-0.5);
\draw[-,thick] (5,-0.5)--(6,-1);
\fill (0,4) circle [radius=2pt];
\fill (0,2) circle [radius=2pt];
\fill (0,0) circle [radius=2pt];
\fill (2,-1) circle [radius=2pt];
\fill (2,5) circle [radius=2pt];
\fill (2,3) circle [radius=2pt];
\fill (2,1) circle [radius=2pt];
\fill (4,4) circle [radius=2pt];
\fill (4,0) circle [radius=2pt];
\fill (6,3) circle [radius=2pt];
\fill (6,1) circle [radius=2pt];
\fill (6,-1) circle [radius=2pt];
\node (eh1) at (-0.5,1) {$\textcolor{red}{\hat{e}_1}$};
\node (eh2) at (5.5,4.5) {$\textcolor{red}{\hat{e}_2}$};
\node (eh3) at (6.5,0) {$\textcolor{red}{\hat{e}_3}$};
\node (eh4) at (3,3) {$\textcolor{red}{\hat{e}_4}$};
\node (eh5) at (3,1) {$\textcolor{red}{\hat{e}_5}$};
\node (eh6) at (3,-1) {$\textcolor{red}{\hat{e}_6}$};
\node (v1) at (-0.5,4) {$v_1$};
\node (v2) at (-0.5,2) {$v_2$};
\node (v3) at (-0.5,0) {$v_3$};
\node (v4) at (2,5.5) {$v_4$};
\node (v5) at (1.5,3) {$v_5$};
\node (v6) at (1.5,1) {$v_6$};
\node (v7) at (2,-1.5) {$v_7$};
\node (v8) at (4,4.5) {$v_8$};
\node (v9) at (4,0.5) {$v_9$};
\node (v10) at (6.5,3) {$v_{10}$};
\node (v11) at (6.5,1) {$v_{11}$};
\node (v12) at (6,-1.5) {$v_{12}$};
\end{tikzpicture}
\caption{A graph with a choice of $\hat{e}_L$'s. The red lines with arrows represent $\hat{e}_L$ ($L=1,\cdots, 6$), 
and the black lines with arrows represent the other edges. The black lines and the vertices
form a spanning tree of the graph. 
}
\label{fig:G_ehL}       
\end{figure}
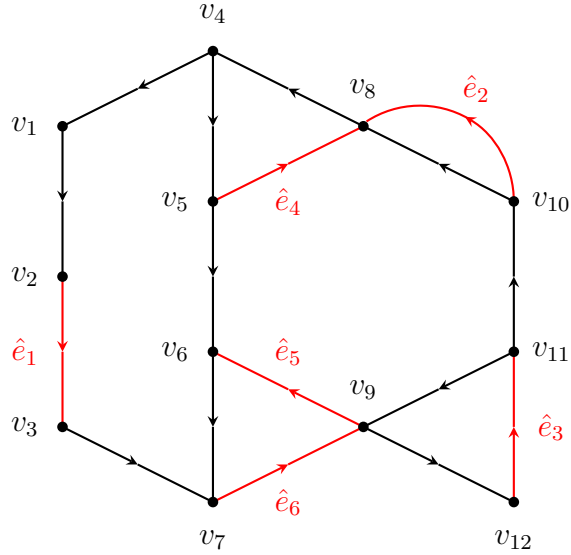

Noting that $A_v^{(bk)}$ acts on the vertex $v$ as identity, the normalization is computed as\footnote{
Similar to the computation in the previous subsection, nonvanishing contribution arises only when the index $b$ of $A_v^{(bk)}$ is the same for all $v$, 
which provides the last equality on the first line of (\ref{cN_comp}).} 
\bea
\langle\mbox{GS}\,s\ket{\mbox{GS}\,s} & =& \cN \bra{s}\prod_{v\in V}A_v\ket{s}=
\cN \left(\frac{1}{n}\right)^{|V|} \bra{s}\left(\prod_{v\in V}\sum_{b=0}^{q-1}A_v^{(bk)}\right)\ket{s}
=\cN\,n^{-|V|}\sum_{b=0}^{q-1}\bra{s}\prod_{v\in V}A_v^{(bk)}\ket{s} \nn \\
& = & \cN\,n^{-|V|}q,
\label{cN_comp}
\eea
which determines $\cN$ as
\be
\cN=\frac{n^{|V|}}{q}.
\label{cN}
\ee

\paragraph{Choice of $\hat{e}_L$'s}
Here, we show that the same set of the independent ground states is obtained irrespective of the choice of $\hat{e}_L$'s. 

When we change the initial choice of one of $\hat{e}_L$'s, say $\hat{e}_{L_0}$, we choose an edge among the edges on the closed path in the graph $T+\{\hat{e}_{L_0}\}$, instead of $\hat{e}_{L_0}$. 
Suppose we pick an edge $e'$ instead of $\hat{e}_{L_0}$. 
Then the graph $T-\{e'\}$, the spanning tree $T$ after the edge $e'$ is removed, splits into two connected tree graphs, which are denoted by $T_1$ and $T_2$. 
We can see that $\prod_{v\in T_1}A_v^{(bk)}$ becomes the product of $X_e^{\pm pbk}$'s with respect to the edges $e=\hat{e}_{L_0}$, $e'$ and some other $\hat{e}_L$'s, 
where $\pm$ in the power is fixed by the orientation. 
$\prod_{v\in T_2}A_v^{(bk)}$ gives essentially the same result, 
since $\prod_{v\in T_2}A_v^{(bk)}=  \left(\prod_{v\in V}A_v^{(bk)}\right)\prod_{v\in T_1}A_v^{(-bk)}=\prod_{v\in T_1}A_v^{(-bk)}$.  

Noting that
\be
\{X_e^{bk}\left|\right. b \in \mathbb{Z}_q\} = \{X_e^{pbk}\left|\right. b \in \mathbb{Z}_q\}
\label{Xeset}
\ee
due to $\gcd(p,q)=1$, we label the state $\ket{s}$ by $a_v$'s and $pb_L$'s:
\be
\ket{s}=\left(\prod_{v\in V}x_v^{a_v}\bX_{L_v}^{a_v}\right)\left(\prod_{L=1}^{B_1}X_{\hat{e}_L}^{pb_Lk}\right)\ket{s=0}.
\label{kets_initial}
\ee
The above result leads to
\be
\left(\prod_{v\in T_1}A_v^{(-b_{L_0}k)}\right)\ket{s}= \left(\prod_{v\in V}x_v^{a_v}\bX_{L_v}^{a_v}\right)\left(\prod_{L\neq L_0}X_{\hat{e}_L}^{pb'_Lk}\right)X_{e'}^{-pb_{L_0}k}\ket{s=0},
\label{kets_change}
\ee
where $b'_L=b_L\pm b_{L_0}$ for the edge $\hat{e}_L$ appearing in the result of $\prod_{v\in T_1}A_v^{(bk)}$, otherwise $b'_L=b_L$.  
In (\ref{kets_change}) $e'$ appears with the label $-pb_{L_0}$ instead of $\hat{e}_{L_0}$, and some other $\hat{e}_L$'s remain with the label changed by $\pm pb_{L_0}$,  
compared to the initial choice (\ref{kets_initial}). The RHS of (\ref{kets_change}) describes the seed state with the choice of $\hat{e}_{L_0}$ changed to $e'$, and the LHS 
shows that it provides the same ground state as (\ref{kets_initial}) as we saw below (\ref{Avbk}). 
Thus, we can say that the ground state is invariant under the change of $\hat{e}_{L_0}$ accompanied with appropriate change 
of the labels $\{b_L\}$. 
Since the number of the labels $\{b_L\}$ does not change before and after the change of the choice, the set of the independent ground states remains the same. 

Let us illustrate the above in the graph in Fig.~\ref{fig:G_ehL}. 
We consider the case $\hat{e}_{L_0}=\hat{e}_5$, and take $e'=e_{48}$ in the closed path on the graph $T+\hat{e}_5$. 
Then, $T-\{e_{48}\}$ splits into the two connected tree graphs: $T_1$ composed by the vertices $v_1,\cdots, v_7$ and the black edges connecting them, 
and $T_2$ composed by $v_8,\cdots, v_{12}$ and the black edges connecting them. 
We have
\be
\prod_{v\in T_1}A_v^{(bk)}=\left(X_{\hat{e}_5}X_{e_{48}}X_{\hat{e}_4}^{-1}X_{\hat{e}_6}^{-1}\right)^{pbk}. 
\ee
Acting $\prod_{v\in T_1}A_v^{(-b_5k)}$ on the initial choice (\ref{kets_initial}) with $B_1=6$ leads to
\be
\left(\prod_{v\in T_1}A_v^{(-b_5k)}\right) \ket{s}=\left(\prod_{v\in V}x_v^{a_v}\bX_{L_v}^{a_v}\right)\left(\prod_{L=1}^3X_{\hat{e}_L}^{pb_Lk}\right)
X_{\hat{e}_4}^{p(b_4+b_5)k}X_{\hat{e}_6}^{p(b_6+b_5)k}X_{e_{48}}^{-pb_5k}\ket{s=0}.
\ee
Thus, the ground state (\ref{GSs}) remains the same under the change of $\hat{e}_5$ to $e_{48}$ 
together with labels changed as $b_5\to -b_5$, $b_4\to b_4+b_5$ and $b_6 \to b_6+b_5$ (mod $q$). 

Since this procedure can be repeated for changes of the other $\hat{e}_L$'s, we can say that the same set of the independent ground states is obtained irrespective of the choice of  $\hat{e}_L$'s. 

The obtained ground states $\ket{\mbox{GS}\,s}$ in (\ref{GSs}) and (\ref{kets_initial}) with $s=0,1,\cdots, (\mbox{GSD})-1$ are eigenstates of the local operator $z_v^k$ (${}^\forall v\in V$) 
and the operator of quasi-local symmetry $Z(C)$ in (\ref{ZC}) for any closed path $C$. Let $C_L$ denote a closed path appearing when $\hat{e}_L$ is added to the spanning tree $T$. 
$\ket{\mbox{GS}\,s}$ is distinguished by the eigenvalues of $z_v^k$'s and $Z(C_L)$'s which measure $a_v$'s and $b_L$'s respectively:
\bea
& & z_v^k\,\ket{\mbox{GS}\,s}=\omega_p^{a_v}\,\ket{\mbox{GS}\,s} ,  \\
 & & Z(C_L)\,\ket{\mbox{GS}\,s}=\omega_q^{pb_L\,(\hat{e}_L|C_L)}\,\ket{\mbox{GS}\,s}.
 \label{ZCL_GSs}
\eea
In order to give a physical interpretation of (\ref{ZCL_GSs}), let us pick an embedding space of the graph in which a simply connected domain bounded by $C_L$ can be defined.\footnote{
Graphs consist of vertices and edges as mentioned in the beginning of section~\ref{Operators}. Since the domain bounded by $C_L$ lies outside the graph, 
we need to mention the embedding space in order to consider magnetic flux penetrating the domain. This is analogous to the global magnetic fluxes penetrating the hole of the torus in the toric code. Incidentally they also distinguish the ground states in the toric code just as how the local magnetic fluxes distinguish the ground states in our models. 
}   
From the analogy to the field theory in appendix~\ref{app:analogy}, $a_v$ represents some degrees of freedom 
of the matter field $\phi$ at the point $v$ on the ground state, 
while $pb_L(\hat{e}_L|C_L)$ is interpreted as magnetic flux penetrating the inside of $C_L$ since the Wilson loop measures magnetic flux penetrating the domain surrounded by the loop. 
This is valid even if $C_L$ is a topologically nontrivial cycle under the setting of the embedding space. 
Note that for any $L'(\neq L)$, $\hat{e}_{L'}$ does not belong to $C_L$, which is seen from the above definition of $C_L$. 
$U(1)$ gauge field on a circle has nontrivial topological structure as briefly summarized in appendix~\ref{app:U(1)}. 
In particular, the nontrivial topological configuration of the gauge field generates magnetic flux as seen in (\ref{Phi}), 
which is analogous to the twist by $X_{\hat{e}_L}^{pb_Lk}$ providing the $\mathbb{Z}_q$ magnetic flux $pb_L$. 

On the other hand, an individual ground state is not invariant under the local transformations $x_v\bX_{L_v}$ and $X_e^k$, 
but mapped to another individual ground state. 

\subsection{Ground states $\ket{\mbox{GS}[\alpha,\beta]}$}
\label{sec:GSab}
Next, we construct eigenstates with respect to the local transformations $x_v\bX_{L_v}$ and $X_e^k$ by taking 
appropriate linear combinations of the ground states $\ket{\mbox{GS}\,s}$ ($s=0,1,\cdots, (\mbox{GSD})-1$). 
Let us introduce operators
\bea
P_v^{[\alpha_v]} & \equiv &\frac{1}{p}\sum_{a=0}^{p-1}\omega_p^{a\alpha_v}x_v^a {\bf X}_{L_v}^a \qquad (\alpha_v\in \mathbb{Z}_p), 
\label{Pv} \\
P_e^{[\beta_e]} & \equiv & \frac{1}{q}\sum_{b=0}^{q-1}\omega_q^{b\beta_e}X_e^{bk}  \qquad (\beta_e\in \mathbb{Z}_q). 
\label{Pe}
\eea
Note that 
\be
P_v^{[\alpha_v]} =A_v P_v^{[\alpha_v]} =\frac{1}{np}\sum_{j=0}^{np-1}\omega_p^{j\alpha_v}x_v^j {\bf X}_{L_v}^j
\label{PvGS}
\ee
holds on the ground states. As the RHSs of (\ref{Pe}) and (\ref{PvGS}) show, they are projection operators on the ground states. 

Acting the operator 
\be
\cP=\cP^{[\alpha,\beta]}\equiv\left( \prod_{v\in V}P_v^{[\alpha_v]} \right)\left(\prod_{L=1}^{B_1}P_{\hat{e}_L}^{[\beta_L]}\right).
\label{calP}
\ee
on $\ket{\mbox{GS}\,0}$  
generates a desirable linear combination of all the ground states with the coefficients being phases: 
\be
\ket{\mbox{GS}[\alpha,\beta]}= \cP\ket{\mbox{GS}\,0}. 
\label{GSab}
\ee
Compared with the ground states $\ket{\mbox{GS}\,s}$ in (\ref{GSs}), 
the labels $a_v$'s and $b_L$'s are converted to $\alpha_v$'s and $\beta_L$'s by the discrete Fourier transformations.

Associated to the ground state (\ref{GSab}), we can regard the graph as an electric circuit in which the `current' $\beta_L$ flows on the line $\hat{e}_L$ to the direction of its orientation. 
Then, the currents on the other lines which are not $\hat{e}_L$'s are determined by the `current conservation' at the vertices.  
The current conservation follows from the relation $A_v^{(bk)}\ket{\mbox{GS}\,s}=\ket{\mbox{GS}\,s}$, namely
\be
\prod_{e\in L_v^+}X_e^{pbk} = \prod_{e\in L_v^-}X_e^{pbk}  \qquad \mbox{on} \quad \ket{\mbox{GS}\,s}
\label{current_conserv}
\ee
for any $v$ and $b\in \mathbb{Z}_q$.
We can see that 
\bea
 & & x_v {\bf X}_{L_v}\ket{\mbox{GS}[\alpha,\beta]}=\omega_p^{-\alpha_v}\ket{\mbox{GS}[\alpha,\beta]},  
 \label{xvXLv_GSab}\\
 & & X_e^k\,\ket{\mbox{GS}[\alpha,\beta]}= \begin{cases} \omega_q^{-\beta_L}\ket{\mbox{GS}[\alpha,\beta]} & (e=\hat{e}_L) \\
\omega_q^{-\bar{\beta}_e}\ket{\mbox{GS}[\alpha,\beta]} & (e\notin\{\hat{e}_1,\cdots,\hat{e}_{B_1}\}), \end{cases}
\label{Xe_GSab}
\eea 
where $\bar{\beta}_e$ represents the current on the edge $e$, a linear combination of $\beta_L$'s determined by the current conservation. 
Note that (\ref{Pe}) can also be written as
\be
P_e^{[\beta_e]} = \frac{1}{q}\sum_{b=0}^{q-1}\omega_q^{pb\beta_e}X_e^{pbk}  \qquad (\beta_e\in \mathbb{Z}_q)
\label{Pe2}
\ee 
since $p$ and $q$ are coprime. 
As an example, for the graph in Fig.~\ref{fig:G_ehL}, $\bar{\beta}_{e}$'s are determined as $\bar{\beta}_{e_{12}}=\bar{\beta}_{e_{14}}=\bar{\beta}_{e_{37}}=\beta_1$, $\bar{\beta}_{e_{67}}=-\beta_1+\beta_6$, 
$\bar{\beta}_{e_{56}}=-\beta_1-\beta_5+\beta_6$, $\bar{\beta}_{e_{45}}=-\beta_1+\beta_4-\beta_5+\beta_6$, $\bar{\beta}_{e_{48}}=\beta_4-\beta_5+\beta_6$, and so on. 
 
By the discrete Fourier transformations of (\ref{kets_initial}) and (\ref{kets_change}), we can see that when the choice of $\hat{e}_{L_0}$ is changed to $e'$ as discussed in the previous subsection, 
the initial ground state $\ket{\mbox{GS}[\alpha,\beta]}$ remains the same form with the current $\beta_{L_0}$ on $\hat{e}_{L_0}$ replaced to 
$\bar{\beta}_{e'}$ on $e'$. 
From a set of the ground states for any one choice of $\hat{e}_L$'s, the ground states for the other choices are derived.

Thus $\ket{\mbox{GS}[\alpha,\beta]}$ is invariant under the local $\mathbb{Z}_p$ and $\mathbb{Z}_q$ transformations (up to phase factors) as in (\ref{xvXLv_GSab}) and (\ref{Xe_GSab}). 
As $A_v$ can be regarded as the Gauss law operator, $x_v^p\bX_{L_v}^p$  corresponds to an operator of `small gauge transformations', i.e., topologically trivial gauge transformations connected to the identity. 
Then, the local $\mathbb{Z}_p$ and $\mathbb{Z}_q$ transformations, which are not generated by $x_v^p\bX_{L_v}^p$, 
can be interpreted as `large gauge transformations', topologically nontrivial gauge transformations not connected to the identity. 
$\ket{\mbox{GS}[\alpha,\beta]}$ is similar to the $\theta$ vacuum in gauge theory when the vacuum has nontrivial topological structure, as seen in (\ref{thetavac}).

On the other hand, $z_v^k$ and $Z(C_L)$ act on (\ref{GSab}) as
\bea
 & & z_v^k\,\ket{\mbox{GS}[\alpha,\beta]}= \ket{\mbox{GS}[\tilde{\alpha},\beta]}, 
 \label{zv_GSab}
 \\
 & & Z(C_L)\,\ket{\mbox{GS}[\alpha,\beta]}= \ket{\mbox{GS}[\alpha,\tilde{\beta}]},
 \label{ZCL_GSab}
 \eea
where 
\be
\tilde{\alpha}_{v'} \equiv\begin{cases} \alpha_{v'} &(v'\neq v) \\
\alpha_v+1 & (v'=v) \end{cases} 
\qquad \mbox{and} \qquad 
\tilde{\beta}_{L'} \equiv \begin{cases} \beta_{L'} & (L'\neq L) \\
 \beta_L+(\hat{e}_L|C_L) & (L'=L). \end{cases}
\ee
From the analogy to the field theory in appendix~\ref{app:analogy}, $\alpha_v$ represents some degrees of freedom of the momentum $\pi$ of the matter field at the point $v$ on the ground state, 
while $\beta_L$ is interpreted as electric flux or current flowing along $\hat{e}_L$ since the Wilson loop creates the unit electric flux along the loop.\footnote{
We can say that the two ground states $\ket{\mbox{GS}\,s}$ and $\ket{\mbox{GS}[\alpha,\beta]}$ are complementary in the sense of quantum theory. 
Namely, they are analogous to the position eigenstate $\ket{x}$ and the momentum eigenstate $\ket{p}$ with the commutation relation of the corresponding operators 
$[\hat{x},\hat{p}]=\mathrm{i}$. 
The relation $\ket{p}=\int dx\,e^{\mathrm{i}px}\ket{x}$ is similar to the ground states related by the discrete Fourier 
transformations. 
The operator $e^{\mathrm{i}a\hat{x}}$, which is analogous to $z_v^k$ or $Z(C_L)$, measures the position when it acts on the position eigenstate, 
whereas it creates a shift of the momentum when it acts on the momentum eigenstate. 
}   

\paragraph{Norm of the ground state $\ket{\mbox{GS}[\alpha,\beta]}$}
Since the operator (\ref{calP}) satisfies $\cP=\cP^\dagger$, $\cP^2=\cP$ and commutes with $A_v$ for any $v$, 
\bea
\lefteqn{\langle\mbox{GS}[\alpha,\beta]\ket{\mbox{GS}[\alpha,\beta]} = \cN \bra{s=0}\left(\prod_{v\in V}A_v\right)\cP\ket{s=0}}
\nn \\
& = & \cN \left(\prod_{e\in E}\bra{0_e}\right)\left\{\prod_{v\in V}\bra{0_v}A_vP_v^{[\alpha_v]}\ket{0_v}\right\} \left(\prod_{L=1}^{B_1}P_{\hat{e}_L}^{[\beta_L]}\right)\left(\prod_{e\in E}\ket{0_e}\right).
\label{normGSab}
\eea
Then, 
\be
\bra{0_v}A_vP_v^{[\alpha_v]}\ket{0_v} = \bra{0_v}A_v\left(P_v^{[\alpha_v]}\right)^\dagger \ket{0_v}
= \frac{1}{np}\sum_{j=0}^{n-1}\sum_{a=0}^{p-1}\omega_p^{-a\alpha_v} {\bf X}_{L_v}^{pj-a} \delta_m(pj, a), 
\label{AP}
\ee
where $\bra{0_v}x_v^{pj-a}\ket{0_v}$ gives $\delta_m(pj, a)$. 
The nonzero contribution comes from $a=0$ and $j=ku$ ($u=0,1,\cdots, q-1$), which leads to
\be
\bra{0_v}A_vP_v^{[\alpha_v]}\ket{0_v} = 
 \frac{1}{np}\sum_{j=0}^{q-1} {\bf X}_{L_v}^{kpj}, 
 \label{APf}
\ee
and then 
\be
\langle\mbox{GS}[\alpha,\beta]\ket{\mbox{GS}[\alpha,\beta]} = \cN \left(\frac{1}{np}\right)^{|V|} 
\left(\prod_{e\in E}\bra{0_e}\right)\left(\prod_{v\in V}\sum_{j=0}^{q-1}{\bf X}_{L_v}^{kpj}\right) \left(\prod_{L=1}^{B_1}P_{\hat{e}_L}^{[\beta_L]}\right)\left(\prod_{e\in E}\ket{0_e}\right).
\ee
In evaluating this, we combine the results 
\be
\bra{0_e}X_e^{kpj}X_e^{-kpj'}\ket{0_e}=\delta_q(j,j')
\label{delta_edge}
\ee
and\footnote{(\ref{omega_edge}) is derived as follows. Using (\ref{Pe2}), 
the LHS becomes 
\[
\frac{1}{q}\sum_{b=0}^{q-1}\omega_q^{bp\beta_L}\bra{0_{\hat{e}_L}} X_{\hat{e}_L}^{kp(j-j'+b)}\ket{0_{\hat{e}_L}}=\frac{1}{q}\sum_{b=0}^{q-1}\omega_q^{bp\beta_L}\delta_n(kp(j-j'+b),0).
\]
The mod $n$ Kronecker delta is nonzero only when $b=j'-j$ mod $q$, which gives the RHS of (\ref{omega_edge}).
} 
\be
\bra{0_{\hat{e}_L}}X_{\hat{e}_L}^{kpj}X_{\hat{e}_L}^{-kpj'}P_{\hat{e}_L}^{[\beta_L]}\ket{0_{\hat{e}_L}}
=\frac{1}{q}\,\omega_q^{\beta_L\,p(j'-j)}. 
\label{omega_edge}
\ee
%
\begin{figure}[H]
\centering
\captionsetup{width=.8\linewidth}
\begin{tikzpicture}
\fill (0,0) circle [radius=2pt];
\fill (2,0) circle [radius=2pt];
\draw [->,thick] (0,0)--(1,0);
\draw [-,thick] (1,0)--(2,0);
\node (v) at (2,0.5) {$v$};
\node (v') at (0,0.5) {$v'$};
\node (j) at (2,-0.5) {$j$};
\node (j') at (0,-0.5) {$j'$};
\node (e) at (1,-0.5) {$e$};
\draw [-,thick] (0,0)--(-0.3,0.3);
\draw [-,thick] (0,0)--(-0.3,-0.3);
\draw [-,thick] (2,0)--(2.3,0.3);
\draw [-,thick] (2,0)--(2.4,0);
\draw [-,thick] (2,0)--(2.3,-0.3);
\end{tikzpicture}
\caption{The indices $j$ and $j'$ in the figure represents the operators $\bX_{L_v}^{kpj}$ and $\bX_{L_{v'}}^{kpj' }$ sitting at the vertices $v$ and $v'$, respectively. 
The edge $e$ connects the vertices. 
These operators yield the operators in the LHS of eq.(\ref{delta_edge}). The result of (\ref{delta_edge}) shows that only the case $j=j'$ (mod $q$) is relevant.}
\label{fig:delta_edge}       
\end{figure}
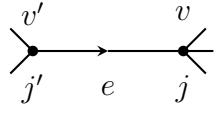
Let us first consider (\ref{delta_edge}) for $e\in E-\{\hat{e}_1,\cdots,\hat{e}_{B_1}\}$. As seen in Fig.~\ref{fig:delta_edge}, The indices $j$ and $j'$ are associated to the vertices 
$v$ and $v'$ at both ends of the edge $e$. For giving nonzero contribution, $j$ and $j'$ should be equal (mod $q$).    
%
From the fact that the graph $T=G-\{\hat{e}_1,\cdots,\hat{e}_{B_1}\}$ is a connected tree graph, all the vertices are connected by the mod $q$ Kronecker delta (\ref{delta_edge}), 
which makes all the indices $j$ equal. Then $\omega_q$-factor in (\ref{omega_edge}) becomes 1. 
Finally we have
\be
\langle\mbox{GS}[\alpha,\beta]\ket{\mbox{GS}[\alpha,\beta]} = \cN \left(\frac{1}{np}\right)^{|V|} \left(\sum_{j=0}^{q-1}1\right)\frac{1}{q^{B_1}}
=\frac{1}{\textrm{GSD}},
\label{normGSabf}
\ee
where we used (\ref{cN}).

\paragraph{Pure-state density matrix}
Next, when $\cP$ in (\ref{calP}) acts on other $\ket{\mbox{GS}\,s}$ with $s\neq 0$, it also gives $\ket{\mbox{GS}[\alpha,\beta]}$ up to some phase factors:
\be
\cP\ket{\mbox{GS}\,s} = e^{\mathrm{i}\theta_s}\ket{\mbox{GS}[\alpha,\beta]},
\ee
which leads to 
\be
\cP \ket{\mbox{GS}\,s} \bra{\mbox{GS}\,s} \cP^\dagger=\ket{\mbox{GS}[\alpha,\beta]}\bra{\mbox{GS}[\alpha,\beta]}
\ee
for any $s$. 

Finally, we find that the desired pure-state density matrix is given by the $\cP$ transformation to the projector (\ref{pi0}),  
$\pi_0=\prod_{v\in V}A_v\prod_{e\in E}B_e=\sum_{s=0}^{(\textrm{GSD})-1}\ket{\mbox{GS}\,s} \bra{\mbox{GS}\,s}$: 
\be
\rho^{[\alpha,\beta]} \equiv  \cP\pi_0\cP^\dagger 
= \sum_{s=0}^{(\textrm{GSD})-1}\cP\ket{\mbox{GS}\,s} \bra{\mbox{GS}\,s}\cP^\dagger 
= (\textrm{GSD})\times \ket{\mbox{GS}[\alpha,\beta]}\bra{\mbox{GS}[\alpha,\beta]}.
\label{rhoab}
\ee
(\ref{normGSabf}) means that $\rho^{[\alpha,\beta]}$ has the correct normalization $\Tr\rho^{[\alpha,\beta]}=1$. 
From the properties of $\cP$, $\rho^{[\alpha,\beta]}$ is simplified as 
\be
\rho^{[\alpha,\beta]}= \left(\prod_{v\in V}A_v\prod_{e\in E}B_e\right) \cP.
\label{rhoab2}
\ee

\section{Entanglement Entropy}
\label{sec:EE}
To better understand the ground states of the Hamiltonian (\ref{H}), we compute their EE. 
As the system is gapped, we expect that the leading order term is proportional to the `area' of the boundary of a bipartition of the system (the {\it area law}), 
which is proven in 
gapped one-dimensional systems~\cite{hastings}. 
Further interesting features are expected in a constant sub-leading term, called {\it topological EE} \cite{kp, zan1, zan2, sb, wen1}, 
which is a speculated signal for a topologically ordered state. 
This is a global term to the EE that is present regardless of the partition of the system.

In this section, we exactly compute the EE (and thus the topological EE) both for the individual ground states $\ket{\mbox{GS}\, s}$ and their linear combinations $\ket{\mbox{GS}[\alpha,\beta]}$, 
with respect to a bipartite separation of the system.  
We first split the total system given by the graph $G$ into the three parts:
\be
G = G_1 + G_2 +E_{12},
\label{division}
\ee
where 
each of $G_1$ and $G_2$ is a connected subgraph, and $E_{12}$ is a set of edges connecting $G_1$ (at the vertices $\nu_1,\cdots, \nu_r$) and 
$G_2$ (at the vertices $\bar{\nu}_1,\cdots,\bar{\nu}_{r'}$). An example of the division (\ref{division}) is depicted in Fig.~\ref{fig:G1G2E12}.
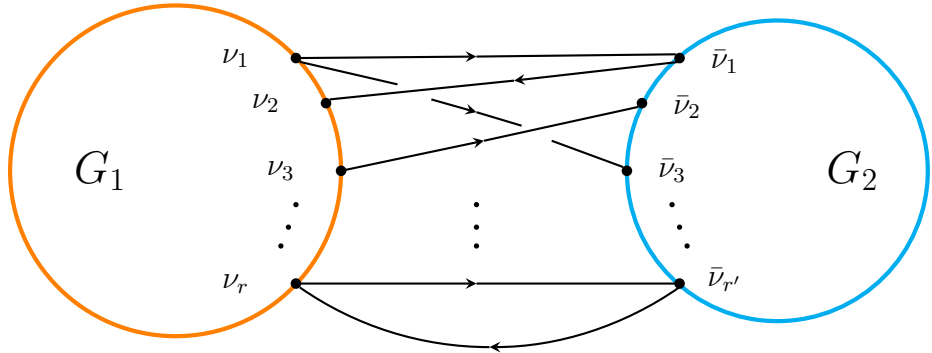
\begin{figure}[h]
\centering
\captionsetup{width=.8\linewidth}
\begin{tikzpicture}
\draw [orange,ultra thick] (-4,0) circle (2.2);
\draw [cyan,ultra thick] (4,0) circle (2);
\fill (-2.4,1.5) circle [radius=2pt];
\fill (-2,0.9) circle [radius=2pt];
\fill (-1.8,0) circle [radius=2pt];
\fill (-2.4,-0.45) circle [radius=1pt];
\fill (-2.5,-0.75) circle [radius=1pt];
\fill (-2.6,-1) circle [radius=1pt];
\fill (-2.4,-1.5) circle [radius=2pt];
\fill (2.7,1.5) circle [radius=2pt];
\fill (2.2,0.9) circle [radius=2pt];
\fill (2,0) circle [radius=2pt];
\fill (2.6,-0.45) circle [radius=1pt];
\fill (2.7,-0.75) circle [radius=1pt];
\fill (2.8,-1) circle [radius=1pt];
\fill (2.7,-1.5) circle [radius=2pt];
\draw [->,thick] (-2.4,1.5)--(0,1.52);
\draw[-,thick] (0,1.52)--(2.65,1.55);
\draw [->,thick] (2.7,1.45)--(0.5,1.2);
\draw[-,thick](0.5,1.2)--(-1.95,0.95);
\draw [->,thick] (-1.8,0)--(0.1,0.4);
\draw[-,thick] (0.1,0.4)--(2.15,0.85);
\draw [-,thick] (-2.35,1.45)--(-1.05,1.15);
\draw [->,thick] (-0.6,0.95)--(0,0.78);
\draw[-,thick] (0,0.78)--(0.6,0.6);
\draw [-,thick] (1,0.4)--(1.95,0.05);
\draw [->,thick] (-2.35,-1.5)--(0,-1.5);
\draw[-,thick] (0,-1.5)--(2.65,-1.5);
\draw [<-,thick] (0.174,-2.345) arc (270:305:4.4cm);
\draw [-,thick] (-2.35,-1.55) arc (235:270:4.4cm);
\fill (0,-0.45) circle [radius=1pt]; 
\fill (0,-0.75) circle [radius=1pt];
\fill (0,-1) circle [radius=1pt];
\node (nu1) at (-3.2,1.5) {$\nu_1$};
\node (nu2) at (-2.8,0.9) {$\nu_2$};
\node (nu3) at (-2.6,0) {$\nu_3$};
\node (nur) at (-3.2,-1.5) {$\nu_r$};
\node (nub1) at (3.3,1.45) {$\bar{\nu}_1$};
\node (nub2) at (2.8,0.85) {$\bar{\nu}_2$};
\node (nub3) at (2.6,0) {$\bar{\nu}_3$};
\node (nubrp) at (3.3,-1.45) {$\bar{\nu}_{r'}$};
\node (G1) at (-5,0) {\Large{$G_1$}};
\node (G2) at (5,0) {\Large{$G_2$}};
\end{tikzpicture}
\caption{An example of the division (\ref{division}). The orange (light blue) circle and its interior represent the region where the connected subgraph $G_1$ ($G_2$) is located. 
Edges and vertices in the interior are suppressed. 
The black lines with arrows are edges belonging to $E_{12}$, 
which connect $G_1$ at the vertices $\nu_1,\nu_2,\cdots,\nu_r$ and $G_2$ at the vertices $\bar{\nu}_1,\bar{\nu}_2,\cdots,\bar{\nu}_{r'}$. }
\label{fig:G1G2E12}       
\end{figure}
$N_v$ and $N_e$ denote the numbers of vertices and edges of $G_1$, and $M_v$ and $M_e$ denote those of $G_2$. $E_{12}$ consists of $f$ edges. 
Then, the first Betti numbers of $G_1$ and $G_2$ are given by
\be
B_1'=N_e-N_v+1 \qquad \mbox{and} \qquad
B_1''=M_e-M_v+1,
\label{bettiG1G2}
\ee
respectively. 
Then the first Betti number of $G$ is 
\be
B_1=|E|-|V|+1=(N_e+M_e+f)-(N_v+M_v)+1 = B_1'+B_1''+f-1.
\label{bettiG}
\ee

For computing the bipartite EE, we take a subsystem $A$ as $G_1$ and trace out the degrees of freedom of the rest $B=G_2+E_{12}$. 
For each $t=1,\cdots, r$, we divide a set of the edges attaching to the vertex $\nu_t$, $L_t (\equiv L_{\nu_t})$,\footnote{In this section, 
we often write the subscripts $\nu_t$ ($t=1, \cdots ,r$) and $\bar{\nu}_{t}$ $(t=1,\cdots, r'$) as $t$ and $\bar{t}$ for notational simplicity. 
Then $\bar{t}$ runs over $\bar{1}, \cdots, \bar{r'}$.
} 
into a set of those belonging to $G_1$, $L'_{t}(=L_{t}\cap G_1)$, 
and a set of those belonging to $E_{12}$, $\Lti_{t}(=L_{t}\cap E_{12})$:
\be
L_{t}=L'_{t}+\Lti_{t}\qquad (t=1,\cdots, r).
\label{Lnut}
\ee
Likewise, for $\bar{t}=\bar{1},\cdots,\bar{r'}$, $L_{\bar{t}} (\equiv L_{\bar{\nu}_{t}})$ is divided into $L'_{\bar{t}}(=L_{\bar{t}}\cap G_2)$ and $\Lti_{\bar{t}}(=L_{\bar{t}}\cap E_{12})$:
\be
L_{\bar{t}}=L'_{\bar{t}}+\Lti_{\bar{t}}    \quad (\bar{t}=\bar{1},\cdots, \bar{r'}).
\label{Lnubart}
\ee
Correspondingly, $\bX_{L_{t}}$ and $\bX_{L_{\bar{t}}}$ are factored as 
\be
\bX_{L_{t}}=\bX_{L'_{t}}\bX_{\Lti_{t}}\quad  (t=1,\cdots, r)
\qquad \mbox{and}\qquad 
\bX_{L_{\bar{t}}}=\bX_{L'_{\bar{t}}}\bX_{\Lti_{\bar{t}}}\qquad  (\bar{t}=\bar{1},\cdots, \bar{r'}).
\label{Xe_factor}
\ee
Then, $E_{12}=\{\Lti_{1},\cdots,\Lti_{r}\}=\{\Lti_{\bar{1}},\cdots,\Lti_{\bar{r'}}\}$. 
For any edge connecting the vertices $\nu_t\in G_1$ and $\bar{\nu}_{t'}\in G_2$, when the edge is incoming to $\nu_t$, it is outgoing from $\bar{\nu}_{t'}$, and vice versa.
Thus,
\be
\prod_{t=1}^r\bX_{\Lti_{t}}=\prod_{\bar{t}=\bar{1}}^{\bar{r'}}\bX_{\Lti_{\bar{t}}}^{-1}
\label{Xe_nu_barnu} 
\ee
holds.

\subsection{Bipartite EE for $\ket{\mbox{GS}\,s}$}
\label{sec:EEGSs}
We start with a pure state described by the density matrix 
\be
\rho_s=\ket{\mbox{GS}\,s}\bra{\mbox{GS}\,s}
\qquad \mbox{with}\qquad s=0,1,\cdots, (\mbox{GSD})-1,
\ee
and compute the reduced density matrix by tracing out the degrees of freedom of $B=G_2+E_{12}$: 
\be
\rho_{s,\,A} = \Tr_B\,\rho_s=\frac{n^{N_v+M_v}}{q}\,\Tr_B\left[\left(\prod_{v\in V}A_v\right)\ket{s}\bra{s}\left(\prod_{v\in V}A_v\right)\right],
\label{rhosA}
\ee
where (\ref{GSs}) and (\ref{cN}) are used. 
On the RHS of 
\be
\prod_{v\in V}A_v=\left(\prod_{v\in G_1-\{\nu_1,\cdots,\nu_r\}}A_v\right)\left(\prod_{t=1}^{r}A_{t}\right)\left(\prod_{v\in G_2}A_v\right),
\ee
the first factor is irrelevant to the trace, whereas the last factor is fully traced out and the trace cyclicity can be applied. 
$\ket{s}$ in (\ref{kets_initial}) can be expressed as a product state
\be
\ket{s} = \prod_{v\in V}\ket{h_v}\prod_{e\in E}\ket{i_e}\qquad \mbox{for some } h_v\in \mathbb{Z}_m,\,i_e\in \mathbb{Z}_n
\ee
which is similarly decomposed.
Then (\ref{rhosA}) reads 
\be
\rho_{s,\,A} = \frac{n^{N_v+M_v}}{q}\,\cA_1\,
\Tr_B\left[\cA_{\rm bdy}\left(\prod_{v\in G_2}A_v\ket{h_v}\bra{h_v}\right)\left(\prod_{e\in B}\ket{i_e}\bra{i_e}\right)\cA_{\rm bdy}^\dagger\right]
\cA_1^\dagger
\label{rhosA2}
\ee
with 
\bea
& & \cA_1 \equiv \left(\prod_{v\in G_1-\{\nu_1,\cdots,\nu_r\}}A_v\right)\left(\prod_{e\in G_1-\{L'_{1},\cdots,L'_{r}\}}\ket{i_e}\right), 
\label{cA1}\\
& & \cA_{\rm bdy}  \equiv \prod_{t=1}^rA_{t}\ket{h_{t}}\ket{i_{L'_{t}}}, \qquad \ket{i_{L'_{t}}}\equiv\prod_{e\in L'_{t}}\ket{i_e}. 
\label{cAbdy}
\eea
Note that $\cA_1$ and $\cA_{\rm bdy}$ act nontrivially on the Hilbert space on $e\in \{L'_{1},\cdots,L'_{r}\}\subset G_1$ and that on $e\in \{\Lti_{1},\cdots,\Lti_{r}\}=E_{12}$, respectively. 

\paragraph{Computation of $\Tr_{G_2}$}
In the computation of $\Tr_B=\Tr_{E_{12}}\Tr_{G_2}$, let us first compute $\Tr_{v\in G_2}$ and then $\Tr_{e\in G_2}$.  

The relevant part of the former is only the second factor in $\Tr_B[\cdots]$ in (\ref{rhosA2}): 
\be
\Tr_{v\in G_2}\left[\prod_{v\in G_2}A_v\ket{h_v}\bra{h_v}\right]=\prod_{v\in G_2}\bra{h_v}A_v\ket{h_v}=\left(\frac{1}{n}\right)^{M_v}\prod_{v\in G_2}\sum_{b=0}^{q-1}\bX_{L_v}^{pbk}.
\label{TrG2-1}
\ee
As is seen in (\ref{Avbk}), only $j=bk$ ($b\in \mathbb{Z}_q$) terms in (\ref{vo}) give nonvanishing contribution. 
Next, $\Tr_{e\in G_2}$ is computed as 
\bea
& & \Tr_{e\in G_2}\left[\left(\prod_{v\in G_2}\sum_{b=0}^{q-1}\bX_{L_v}^{pbk}\right)\left(\prod_{e\in G_2}\ket{i_e}\bra{i_e}\right)\right] 
=\left(\prod_{e\in G_2}\bra{i_e}\right)\left(\prod_{v\in G_2}\sum_{b=0}^{q-1}\bX_{L_v}^{pbk}\right)\left(\prod_{e\in G_2}\ket{i_e}\right) \nn \\
& & = \sum_{b=0}^{q-1}\,\prod_{\bar{t}=\bar{1}}^{\bar{r'}}\bX_{\Lti_{\bar{t}}}^{pbk}=\sum_{b=0}^{q-1}\,\prod_{t=1}^{r}\bX_{\Lti_{t}}^{pbk},
\label{TrG2-2}
\eea
where (\ref{Xe_nu_barnu}) and the change of the summation index $b\to q-b$ were used at the last equality. 

Now we find 
\be
\rho_{s,\,A} = \frac{n^{N_v}}{q}\,\cA_1\,\Tr_{E_{12}}\left[\cA_{\rm bdy}\left(\sum_{b=0}^{q-1}\,\prod_{t=1}^{r}\bX_{\Lti_{t}}^{pbk}\right)\left(\prod_{e\in E_{12}}\ket{i_e}\bra{i_e}\right)\cA_{\rm bdy}^\dagger\right]
\cA_1^\dagger.
\label{rhosA3}
\ee

\paragraph{Computation of $\Tr_{E_{12}}$}
From (\ref{Lnut}), $A_{t}(\equiv A_{\nu_t})$ can be expressed as $A_{t}=\frac{1}{n}\sum_{j=0}^{n-1}x^{pj}_{t}\bX_{L'_{t}}^{pj}\bX_{\Lti_{t}}^{pj}$. 
$\Tr_{E_{12}}[\cdots]$ in (\ref{rhosA3}) becomes 
\bea
 & & (\mbox{$\Tr_{E_{12}}[\cdots]$ in (\ref{rhosA3})}) = \sum_{b=0}^{q-1}\prod_{t=1}^r\left\{\left(\frac{1}{n}\right)^2\sum_{j,j'=0}^{n-1}x_{t}^{pj}\ket{h_{t}}\,\bX_{L'_{t}}^{pj}\ket{i_{L'_{t}}}\right.\nn\\
 & & \left.\times \Tr_{\Lti_{t}}\left[\bX_{\Lti_{t}}^{pj+pbk}\left(\prod_{e\in \Lti_{t}}\ket{i_e}\bra{i_e}\right)\bX_{\Lti_{t}}^{-pj'}\right]
 \bra{i_{L'_{t}}}\bX_{L'_{t}}^{-pj'}\,\bra{h_{t}}x_{t}^{-pj'}\right\},
 \label{TrE12-1}
\eea
in which $\Tr_{\Lti_{t}}[\cdots]$ gives $\delta_n\left(p(-j'+j+bk),0\right)$. 
From (\ref{gcd_kti_pti}), the Kronecker delta means that $\pti (-j'+j+bk) = 0$ mod $\kti q$. Note that $\gcd(p,q)=\gcd(\xi\pti,q)=1$ is equivalent to $\gcd(\xi,q)=1$ and $\gcd(\pti,q)=1$. Combining this and (\ref{gcd_kti_pti}), we find 
$\gcd(\pti,\kti q)=1$. Thus, $j'$ giving nonzero contribution is $j'=j+bk$ mod $\kti q$, i.e., 
\be
j'=j+bk-\kti qu \qquad (u\in \mathbb{Z}_\xi),
\ee
which leads to
\be
 (\mbox{$\Tr_{E_{12}}[\cdots]$ in (\ref{rhosA3})}) = \left(\sum_{b=0}^{q-1}\,\prod_{t=1}^r\bX_{L'_{t}}^{-pbk}\right)\prod_{t=1}^r\left\{\frac{\xi}{n^2}\left(\sum_{j=0}^{n-1}P_{t,\,j}P_{L'_{t},\,j}\right)Q_{t}\right\}.
 \label{TrE12-2}
\ee
Here $P_{t,\,j}$, $P_{L'_{t},\,j} $ and $Q_{t}$ are projection operators defined by
\be
P_{t,\,j} \equiv x_{t}^{pj}\ket{h_{t}}\bra{h_{t}}x_{t}^{-pj}, \qquad
P_{L'_{t},\,j} \equiv \bX_{L'_{t}}^{pj}\ket{i_{L'_{t}}}\bra{i_{L'_{t}}}\bX_{L'_{t}}^{-pj}, \qquad
Q_{t} \equiv \frac{1}{\xi}\sum_{u=0}^{\xi-1}x_{t}^{p\kti qu}.
\label{proj1}
\ee
In deriving (\ref{TrE12-2}), $pbk=mb$ and $p\kti qu=\pti kq u=\pti nu$ were used. 

Since it can be seen that the property
\be
\left(P_{t,\,j}P_{L'_{t},\,j}\right)\left(P_{t,\,j'}P_{L'_{t},\,j'}\right)=\delta_{j,j'}\left(P_{t,\,j}P_{L'_{t},\,j}\right)
\ee
holds,
we introduce more projection operators as
\be
P_t\equiv \sum_{j=0}^{n-1}P_{t,\,j}P_{L'_{t},\,j}, \qquad 
Q'\equiv \frac{1}{q}\sum_{b=0}^{q-1}\,\prod_{t=1}^r\bX_{L'_{t}}^{-pbk},
\label{proj2}
\ee
and obtain 
\be
 (\mbox{$\Tr_{E_{12}}[\cdots]$ in (\ref{rhosA3})}) =\left(\frac{\xi}{n^2}\right)^r q\,Q'\left(\prod_{t=1}^r P_tQ_{t}\right).
 \label{TrE12f}
\ee 
Plugging (\ref{TrE12f}) to (\ref{rhosA3}), we end up with
\be
\rho_{s,\,A}=n^{N_v-2r}\xi^r\,\cA_1\,Q'\left(\prod_{t=1}^r P_tQ_{t}\right)\cA_1^\dagger.
\label{rhosAf}
\ee

Notice that the projection operators $Q'$, $P_t$ and $Q_{t}$ mutually commute. 
$P_tx_{t}^{p\kti qu}=x_{t}^{p\kti qu}P_t$ and $P_t\bX_{L'_{t}}^{-pbk}= \bX_{L'_{t}}^{-pbk}P_t$ are verified by shifting $j$ in the sum in the definition of $P_t$ (\ref{proj2}) as 
$j \to j+\kti qu$ and $j\to j-bk$, respectively.  This leads to $P_tQ_{t}=Q_{t}P_t$ and $P_tQ'=Q'P_t$. 

\paragraph{Bipartite EE}
A similar computation of (\ref{TrG2-1}) and (\ref{TrG2-2}) gives 
\be
\cA_1^\dagger \cA_1=n^{-N_v+r}\sum_{b=0}^{q-1}\prod_{t=1}^r\bX_{L'_{t}}^{-pbk}=n^{-N_v+r}q\,Q'.
\label{cA1dagger_cA1}
\ee
Then we find 
\be
\rho_{s,\,A}^2=\tilde{n}^{-r}q\,\rho_{s,\,A}
\ee
with (\ref{ntilde}). 
It means ${\rm Spec}\,\rho_{s,\,A}=\left\{\tilde{n}^{-r}q ,0\right\}$. 
As it should be from $\Tr \rho_s=1$, we can directly check $\Tr\rho_{s,\,A}=1$ from the expression (\ref{rhosAf}). 
Thus, it is seen that $\rho_{s,\,A}$ has the eigenvalue $\tilde{n}^{-r}q$ with multiplicity $\tilde{n}^rq^{-1}$. 

Finally, the bipartite EE is obtained as 
\bea
S_{s,\,A} & = & -\Tr\left[\rho_{s,\,A}\log_2\rho_{s,\,A}\right]=-\left\{\left(\tilde{n}^{-r}q\right)\log_2\left(\tilde{n}^{-r}q\right)\right\}\times \tilde{n}^rq^{-1} \nn \\
 & = & \left(\log_2\tilde{n}\right)r-\log_2 q.
\label{SsA}
\eea 
The result (\ref{SsA}) is independent of $a_v$'s and $b_L$'s which labels the state $\ket{s}$. Namely, all the individual ground state $\ket{\mbox{GS}\,s}$ gives the same EE. 
As $r$ grows the linear term of $r$ dominates, which shows that contribution around the boundary between the subsystems becomes dominant in the EE. 
Namely the EE obeys the area law. 
On the other hand, the constant term $-\log_2 q$ characterizes a global feature of the entanglement of the ground state, which is called the topological EE~\cite{kp}.\footnote{
In~\cite{kp,wen1}, prescriptions are presented to obtain the constant contribution eliminating the short range effects, 
in which it is not necessary to identify the linear term to be subtracted from the whole expression.
} 
The topological EE is denoted by $-\gamma$. Here we have
\be
\gamma=\log_2 q
\label{TEE}
\ee
and the total quantum dimension is $D=q$. 
For $q=1$ ($m$ is an integer multiple of $n$: $m=pn$), the topological EE vanishes. 

For $r\geq 2$, (\ref{SsA}) vanishes only when $\kti=q=1$ (i.e., $m$ is an integer multiple of $n^2$: $m=\pti n^2$). 
Then, 
the vertex operator $A_v$ reduces to a strictly local operator (nontrivially acting only on the vertex $v$): $A_v=\frac{1}{\xi}\sum_{j=0}^{\xi-1}x_v^{\xi\pti j}1_{L_v}$. 
Since the $A_v$ does not generate entanglement, the ground state 
(\ref{GSs}) becomes a product state.

The ground states $\ket{\mbox{GS}\, s}$ have definite magnetic flux for each independent closed path on the graph. 
This corresponds to the basis state which maximizes the negative of the topological EE, $\gamma$, according to~\cite{asvin,wang1}. 
However, in the next subsection we will see that it does not always mean minimizing the whole EE, which is the case in the toric code~\cite{asvin,wang1}.

\subsection{Bipartite EE for $\rho^{[\alpha,\beta]}$}
\label{sec:EEGSab}
In the computation of the EE for the density matrix (\ref{rhoab2}), 
let us consider the case that among the $\hat{e}_L$'s, the first $B_1'$, $\{\hat{e}_1,\cdots,\hat{e}_{B_1'}\}$, are in $G_1$, 
the next $B_1''$, $\{\hat{e}_{B_1'+1},\cdots, \hat{e}_{B_1'+B_1''}\}$, in $G_2$, and the rest, \{$\hat{e}_{B_1'+B_1''+1},\cdots, \hat{e}_{B_1}\}$, in $E_{12}$. 
From (\ref{bettiG}), we see that $(f-1)$ of the $f$ edges of $E_{12}$ are $\hat{e}_L$'s. 
Notice that any choice of $\hat{e}_L$'s can be reduced to the case as discussed in section~\ref{sec:GSab}.  

The reduced density matrix reads 
\bea
\rho_A^{[\alpha,\beta]} & = & \Tr_B \,\rho^{[\alpha,\beta]} =
\Tr_B\left[\left(\prod_vA_vP_v^{[\alpha_v]}\right)\left(\prod_eB_e\right)\left(\prod_{L=1}^{B_1}P_{\hat{e}_L}^{[\beta_L]}\right)\right] \nn \\
& = & \left(\prod_{v\in G_1-\{\nu_1,\cdots,\nu_r\}}A_vP_v^{[\alpha_v]}\right)\left(\prod_{e\in G_1}B_e\right)\left(\prod_{L=1}^{B_1'}P_{\hat{e}_L}^{[\beta_L]}\right) \nn \\
& & \times \Tr_B
\left[\left(\prod_{t=1}^rA_{t}P_{t}^{[\alpha_{t}]}\right)\left(\prod_{v\in G_2}A_vP_v^{[\alpha_v]}\right)
\left(\prod_{e\in B}B_e\right)\left(\prod_{L=B_1'+1}^{B_1}P_{\hat{e}_L}^{[\beta_L]}\right)\right].
\label{rhoabA}
\eea
Here it is easy to see that for $\left(\prod_{e\in B}B_e\right)$ only $\frac{1}{k}B_e^{(0)}$ in $B_e$ (\ref{eo}) gives nonzero contribution. So we may replace $\left(\prod_{e\in B}B_e\right)$
with the factor $k^{-M_e-f}$. The last line of (\ref{rhoabA}) becomes
\bea
(\textrm{last line of (\ref{rhoabA})}) & = &k^{-M_e-f} \,\Tr_{E_{12}}\left\{\left(\prod_{t=1}^rA_{t}P_{t}^{[\alpha_{t}]}\right)\left(\prod_{L=B_1'+B_1''+1}^{B_1}P_{\hat{e}_L}^{[\beta_L]}\right)\right. \nn \\
& & \left. \times \Tr_{G_2}\left[\left(\prod_{v\in G_2}A_vP_v^{[\alpha_v]}\right)\left(\prod_{L=B_1'+1}^{B_1'+B_1''}P_{\hat{e}_L}^{[\beta_L]}\right)\right]\right\}. 
\label{TrB}
\eea

\paragraph{Computation of $\Tr_{G_2}$}
For the computation of $\Tr_{G_2}$ in (\ref{TrB}), we first evaluate $\Tr_{v\in G_2}$ and then $\Tr_{e\in G_2}$. 

In a similar manner to (\ref{APf}),
\be
\Tr_v\left(A_vP_v^{[\alpha_v]}\right)= \frac{1}{q}\sum_{j=0}^{q-1}{\bf X}_{L_v}^{kpj},
\ee
which leads to
\be
(\mbox{$\Tr_{G_2}[\cdots]$ in (\ref{TrB})})=
q^{-M_v}\,\Tr_{e\in G_2}\left[\prod_{v\in G_2}\left(\sum_{j=0}^{q-1}{\bf X}_{L_v}^{kpj}\right)\cdot
\left(\prod_{L=B_1'+1}^{B_1'+B_1''}P_{\hat{e}_L}^{[\beta_L]}\right)\right].
\ee
Similar to (\ref{delta_edge}) and (\ref{omega_edge}), 
\bea
 & & \Tr_e\left(X_e^{kpj}X_e^{-kpj'}\right)=n\delta_q(j,j') ,
\label{delta_edgeTr}
\\
 & & \Tr_{\hat{e}_L}\left(X_{\hat{e}_L}^{kpj}X_{\hat{e}_L}^{-kpj'} P_{\hat{e}_L}^{[\beta_L]}\right)=\frac{n}{q}\,\omega_q^{\beta_L\,p(j'-j)}.
\label{omega_edgeTr}
\eea

Since the graph $G_2-\{\hat{e}_{B_1'+1},\cdots,\hat{e}_{B_1'+B_1''}\}$ becomes a connected tree graph, 
all the $j$-indices become the same due to the mod $q$ Kronecker delta (\ref{delta_edgeTr}) from each edge, and the $\omega_q$-factors from (\ref{omega_edgeTr}) all become 1. 
Then we find 
\be
(\mbox{$\Tr_{G_2}[\cdots]$ in (\ref{TrB})})=n^{M_e}q^{-M_v-B_1''}\, \sum_{j=0}^{q-1}\,\prod_{t=1}^{r'}{\bf X}_{\Lti_{\bar{t}}}^{kpj},
\label{TrG2}
\ee
where (\ref{Lnubart}) and (\ref{Xe_factor}) are used. 

Plugging (\ref{TrG2}) to (\ref{TrB}) we have
\be
(\textrm{last line of (\ref{rhoabA})}) =  k^{-f}q^{-1}\,\Tr_{E_{12}}\left\{\left(\prod_{t=1}^rA_{t}P_{t}^{[\alpha_{t}]}\right)\left(\prod_{L=B_1'+B_1''+1}^{B_1}P_{\hat{e}_L}^{[\beta_L]}\right)
\sum_{j=0}^{q-1}\,\prod_{t=1}^{r}{\bf X}_{\Lti_{t}}^{-kpj}\right\},
\label{TrB2}
\ee
after (\ref{bettiG1G2}) and (\ref{Xe_nu_barnu}) are used. 

\paragraph{Computation of $\Tr_{E_{12}}$}
Using (\ref{PvGS}) and (\ref{Xe_factor}), 
we express (\ref{TrB2}) as 
\bea
(\textrm{last line of (\ref{rhoabA})}) & = &  k^{-f}q^{-1}(np)^{-r}\sum_{j'=0}^{q-1}\sum_{j_1,\cdots,j_r=0}^{np-1} \left(\prod_{t=1}^r\omega_p^{j_t\alpha_{t}}x_{t}^{j_t}\,{\bf X}_{L'_{t}}^{j_t}\right) \nn \\
& & \times \Tr_{E_{12}}\left\{\left(\prod_{t=1}^r{\bf X}_{\Lti_{t}}^{j_t-kpj'}\right)
\left(\prod_{L=B_1'+B_1''+1}^{B_1}P_{\hat{e}_L}^{[\beta_L]}\right)\right\}.
\label{TrB3}
\eea
Note that $j_t$ is associated to the vertex $\nu_t$ and $j'$ is associated to the vertices $\bar{\nu}_1,\cdots,\bar{\nu}_{r'}$. 

Computation of the trace on each edge goes as 
\be
\Tr_e\left(X_e^{-j_t}X_e^{kpj'}\right)=n\delta_n(j_t,kpj'),
\label{TreXX}
\ee
from which $j_t$ giving nonzero contribution is 
\be
j_t=kpj'+kqu \qquad (\mbox{mod $np$})
\label{jtTrE_delta}
\ee
with $j'=0,1,\cdots, q-1$ and $u=0,1,\cdots, p-1$. 
Also, 
\be
\Tr_{\hat{e}_L}\left(X_{\hat{e}_L}^{-j_t}X_{\hat{e}_L}^{kpj'}P_{\hat{e}_L}^{[\beta_L]}\right) 
=\frac{1}{q}\sum_{b=0}^{q-1}\omega_q^{b\beta_L}\,\Tr_{\hat{e}_L}X_{\hat{e}_L}^{-j_t+kpj'+bk} 
=\frac{n}{q}\sum_{b=0}^{q-1}\omega_q^{b\beta_L}\,\delta_n(j_t,k(b+pj')).
\ee
Here, $j_t$ giving nonzero contribution is 
\be
j_t=k\jti_t \qquad (\jti_t=0,1,\cdots, pq-1),
\label{JtTrE_omega}
\ee
and then we find
\be
\Tr_{\hat{e}_L}\left(X_{\hat{e}_L}^{-j_t}X_{\hat{e}_L}^{kpj'}P_{\hat{e}_L}^{[\beta_L]}\right) 
=\frac{n}{q}\,\omega_q^{\beta_L\,(\jti_t-pj')}.
\label{TreXXP}
\ee

As mentioned above, $(f-1)$ of the $f$ edges in $E_{12}$ are $\hat{e}_L$'s. We assume that the only one edge in $E_{12}$ which is not $\hat{e}_L$ attaches to the vertex $\nu_1$. 
It does not lose generality, since this situation can be always realized by appropriately renaming the vertices $\nu_1,\cdots, \nu_r$. 

Plugging (\ref{TreXX})-(\ref{TreXXP}) to (\ref{TrB3}) leads to
\bea
(\textrm{last line of (\ref{rhoabA})}) & = &  (np)^{-r}\sum_{j'=0}^{q-1}\sum_{u=0}^{p-1}\omega_p^{kq\alpha_{1}u}x_{1}^{kqu}\,{\bf X}_{L'_{1}}^{kpj'} \nn \\
 & & \times \sum_{\jti_2,\cdots,\jti_r=0}^{pq-1}\left(\prod_{t=2}^r\omega_p^{k\alpha_{t}\jti_t}\omega_q^{\beta_t'\,(\jti_t-pj')}\,x_{t}^{k\jti_t}\,{\bf X}_{L'_{t}}^{k\jti_t}\right),
 \label{TrB4}
 \eea
where $\beta_t'$ is the sum of the currents $\beta_L$'s flowing from the vertex $\nu_t$ (with the index $j_t$) to the vertices $\bar{\nu}_1, \cdots, \bar{\nu}_{r'}$ (with $j'$). 
For example, in case that the vertex $\nu_t$ is attached to three $\hat{e}_L$'s in $E_{12}$, say $\hat{e}_{L_1}$, $\hat{e}_{L_2}$ and $\hat{e}_{L_3}$, as in Fig,~\ref{fig:betat'},
we have $\beta_t'=\beta_{L_1}+\beta_{L_2}+\beta_{L_3}$. 
%
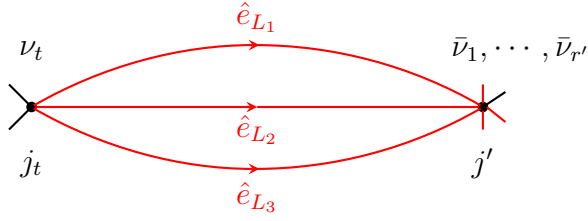
\begin{figure}[H]
\centering
\captionsetup{width=.8\linewidth}
\begin{tikzpicture}
\fill (0,0) circle [radius=2pt];
\fill (6,0) circle [radius=2pt];
\draw [->,thick,red] (0.05,0)--(3,0);
\draw [-,thick,red] (3,0)--(5.95,0);
\draw [->,thick,red] (0.03,0.02) arc (120:90:6cm);
\draw [-,thick,red] (5.97,0.02) arc (60:90:6cm);
\draw [->,thick,red] (0.03,-0.02) arc (240:270:6cm);
\draw [-,thick,red] (5.97,-0.02) arc (-60:-90:6cm);
\node (nut) at (0,0.75) {$\nu_t$};
\node (nubar) at (6.5,0.75) {$\bar{\nu}_1,\cdots,\bar{\nu}_{r'}$};
\node (jt) at (0,-0.75) {$j_t$};
\node (j') at (6,-0.75) {$j'$};
\node (e1) at (3,1.2) {\textcolor{red}{$\hat{e}_{L_1}$}};
\node (e2) at (3,-0.3) {\textcolor{red}{$\hat{e}_{L_2}$}};
\node (e3) at (3,-1.2) {\textcolor{red}{$\hat{e}_{L_3}$}};
\draw [-,thick] (0,0)--(-0.3,0.3);
\draw [-,thick] (0,0)--(-0.3,-0.3);
\draw [-,thick,red] (6,0.05)--(6,0.3);
\draw [-,thick] (6,0)--(6.3,0.2);
\draw [-,thick,red] (6.05,0)--(6.3,-0.2);
\draw [-,thick,red] (6,-0.05)--(6,-0.3);
\end{tikzpicture}
\caption{The index $j_t$ is associated to the vertex $\nu_t$, which is attached to the three edges $\hat{e}_{L_1}$, $\hat{e}_{L_2}$ and $\hat{e}_{L_3}$ in $E_{12}$. 
The other ends of the edges are either of the vertices $\bar{\nu}_1,\cdots,\bar{\nu}_{r'}$ which are depicted by a single dot, because they are endowed with the common index $j'$ 
as a result of the computation of $\Tr_{G_2}$.
In this case, $\beta_t'$ in (\ref{TrB4}) is given by $\beta_t'=\beta_{L_1}+\beta_{L_2}+\beta_{L_3}$.}
\label{fig:betat'}       
\end{figure}

We now define projection operators as 
\bea
 & & Q_{1,p} \equiv  \frac{1}{p}\sum_{u=0}^{p-1}\omega_p^{kq\alpha_{1}u} x_{1}^{kqu}, \qquad 
Q_{1,q}  \equiv \frac{1}{q}\sum_{j'=0}^{q-1}\omega_q^{-pj'(\sum_{t=2}^r\beta_t')}\,{\bf X}_{L_{1}'}^{kpj'}, \nn \\
 & & \tilde{Q}_t \equiv \frac{1}{pq}\sum_{\jti=0}^{pq-1} \omega_p^{k\alpha_{t}\jti}\omega_q^{\beta_t'\jti}\,x_{t}^{k\jti}\,{\bf X}_{L'_{t}}^{k\jti}
\eea
for $t=2,\cdots,r$, and express (\ref{TrB4}) as 
\be
(\textrm{last line of (\ref{rhoabA})}) =k^{-r}Q_{1,p}Q_{1,q}\prod_{t=2}^r\tilde{Q}_t.
\label{TrB5}
\ee
Finally, (\ref{rhoabA}) becomes
\be
\rho_A^{[\alpha,\beta]}
=k^{-r}\left(\prod_{v\in G_1-\{\nu_1,\cdots,\nu_r\}}A_vP_v^{[\alpha_v]}\right)\left(\prod_{e\in G_1}B_e\right)\left(\prod_{L=1}^{B_1'}P_{\hat{e}_L}^{[\beta_L]}\right) 
Q_{1,p}Q_{1,q}\prod_{t=2}^r\tilde{Q}_t.
\label{rhoabAf}
\ee

\paragraph{Result of EE} 
It is easy to see that 
\be
\textrm{Spec}\left(\rho_A^{[\alpha,\beta]}\right)=\{k^{-r}, 0\},
\ee
because the RHS of (\ref{rhoabAf}) is the product of commuting projectors except the factor $k^{-r}$. 
It can be directly checked that $\Tr \rho_A^{[\alpha,\beta]}=1$ holds as it should be from $\Tr\rho^{[\alpha,\beta]}=1$.  
This shows that the reduced density matrix has the eigenvalue $k^{-r}$ with the multiplicity $k^r$. 

The bipartite EE is found as
\be
S_A^{[\alpha,\beta]} = -\Tr\left(\rho_A^{[\alpha,\beta]}\log_2 \rho_A^{[\alpha,\beta]}\right)
=-\left(k^{-r}\log_2k^{-r}\right)\times k^r = (\log_2 k)\,r.
\label{EEabf}
\ee
This is proportional to $r$ (the `area' of the boundary), which exhibits the area law. 
The result is independent of the choice of $\alpha_v$'s or $\beta_L$'s. 
There is no constant term, namely the topological EE vanishes. 

Comparing to the result for the individual ground state $\ket{\mbox{GS}\,s}$ (\ref{SsA}),
we can see that when $\xi=1$ (i.e., $k$ and $p$ are coprime),
\be 
S_A^{[\alpha,\beta]} \leq S_{s,A}
\label{EEabs}
\ee
always holds. 
(\ref{EEabs}) is equivalent to $q^{r-1}\geq 1$, which is valid for any positive integers $q$ and $r$. 
When $\xi \neq 1$, (\ref{EEabs}) is equivalent to 
\be
q\leq \left(\frac{q}{\xi}\right)^r,
\label{EEabs2}
\ee
which holds when $q>\xi$ for $r$ large. 
On the other hand, $S_{s,\,A}$ is smaller than $S_A^{[\alpha,\beta]}$ when $\xi>q$ for any $r$. 

Note that in the case $S_A^{[\alpha,\beta]} < S_{s,A}$ the basis state maximizing the minus of the topological EE does not minimize the EE due to the contribution 
from the leading term proportional to $r$. 
In \cite{asvin}, since the leading term of the EE is common among the bases of the degenerate ground states, 
the basis states which maximize the negative of the topological EE are called the {\it minimum entropy states}. 
However this does not always hold here, because the basis change affects the leading term as well as the constant term of the EE. 
For $q>\xi$ this can be seen as a distinguishing feature of our model from the toric code.

\section{Excited states}
\label{sec:Estates}

In this section we obtain the first and second excited states of the model. 
There are anyon-like excitations among them, and their relevance to the obtained topological EE is discussed. 
Further we discuss the statistics of the obtained anyons for exchange processes that are peculiar to graphs. 

Excited states of the models governed by the Hamiltonian in (\ref{H}) appear when at least one of the edge or vertex operators ($B_e$ or $A_v$) assumes the zero-eigenvalue. 
Recall that each of these operators has the eigenvalues 0 and 1, since they are projection operators. 
In particular when some edge operators (vertex operators) take the zero-eigenvalues we will denote them as {\it edge excitations} ({\it vertex excitations}). 
When both edge and vertex operators take zero-eigenvalues we end up with an example of a {\it combined excitation}. 
Henceforth we use the phrase, `the edge or vertex operators are {\it excited}', when they assume the zero-eigenvalues.  

In what follows we will show that every edge operator can be excited independently, or in other words all the edge excitations are {\it isolated}. 
On the other hand, some of the vertex excitations are isolated and the remaining can only be excited in {\it pairs}, that is they are {\it deconfined} and there is no energy cost in moving them around. This can be contrasted with the situation in the abelian quantum double models where all the excitations are deconfined.

In sections~\ref{sec:eExcitations} and \ref{sec:vExcitations}, we discuss excitations on the ground states $\ket{\mbox{GS}\,s}$. 
Excitations on the ground states $\ket{\mbox{GS}[\alpha,\beta]}$ are similarly constructed.    

\subsection{Edge excitations}
\label{sec:eExcitations}

The edge operator (\ref{eo}) having the zero-eigenvalue implies that one of its orthogonal complements, $B_e^{[\alpha]}$ in (\ref{orthoBe}) with $\alpha \neq 0$ mod $k$, has the eigenvalue 1. 
To check if a single edge operator on $e'$ is excited we follow the computations of the GSD in section~\ref{sec:GSD} to evaluate
\be
\Tr_\mathcal{H}\left(B_{e'}^{\left[\alpha\right]}\prod\limits_{v\in V} A_v\prod\limits_{e\in E-\{ e' \}}B_e \right) = p^{|V|}q^{B_1} = \mbox{GSD}\neq 0,
\ee
which implies the existence of isolated edge excitations for all values of $\alpha \in \{1, 2,\cdots, k-1\}$. This exhausts the elementary edge excitations of the theory meaning that a pair of edge excitations has to be composed of two isolated edge excitations.

\paragraph{Excited states $X_e^\beta\ket{\mbox{GS}\,s}$:}
To obtain these isolated edge excitations, let us first pick the state $X_e^\beta\ket{\mbox{GS}\, s}$ with $\beta\in \{1,2,\cdots,n-1\}$. 
The ground state $\ket{\mbox{GS}\,s}$ is given by (\ref{GSs}) and (\ref{kets_initial}). 
Since $X_e^k$ is a local symmetry mapping the ground state to some other ground state, 
the above state for any $\beta$ reduces to the form $X_e^\beta\ket{\mbox{GS}\,s'}$ with $\beta\in \{1,2,\cdots,k-1\}$.
Hence we may consider the case $\beta\in\{1,2,\cdots,k-1\}$ without loss of generality. 

From (\ref{xz_XZ}) we can see that 
\be
B_e^{(j)}X_e^\beta=\omega_k^{\beta j} X_e^\beta B_e^{(j)}
\ee
and thus 
\be
B_eX_e^\beta = X_e^\beta B_e^{[\beta]}. 
\label{BeXe}
\ee
In addition, since $\ket{s}$ is an eigenstate of $B_e$ with the eigenvalue 1, 
\be
B_e^{[\alpha]}\ket{s}=0 \quad \mbox{for $\alpha\neq 0$ mod $k$}.
\label{Be_s}
\ee 
(\ref{BeXe}) and (\ref{Be_s}) lead to
\be
B_eX_e^\beta\ket{\mbox{GS}\,s}=0 \quad \mbox{for $\beta\in\{1,2,\cdots,k-1\}$},
\ee
which implies that $X_e^\beta\ket{\mbox{GS}\,s}$ are first excited states with the energy $E_0+1$. 
Here, $E_0=-|V|-|E|$ is the ground state energy.  

We can also apply the operators $x_v^\beta$ ($\beta\in\{1,2,\cdots,k-1\}$) on the vertices to excite all the edge operators corresponding to the edges attached to the vertex $v$. 
As discussed in section~\ref{sec:H}, any power of $x_v$ reduces to the above $x_v^\beta$ up to the multiplications of the local symmetry operator $\left(x_v\bX_{L_v}\right)^{kq}=x_v^{kq}$. 
However, $x_v^\beta$ is not an independent excitation but a collection of the isolated excitations on the edges attached to the vertex $v$. This follows from 
\be
x_v^\beta =  \left(x_v^\beta \bX_{L_v}^\beta\right) \bX_{L_v}^{-\beta},
\ee
where the factor in the parentheses generates a local symmetry. 
Thus, $x_v^\beta\ket{\mbox{GS}\, s}=\bX_{L_v}^{-\beta}\ket{\mbox{GS}\,s'}$ with $\ket{s'} =  \left(x_v^\beta \bX_{L_v}^\beta\right)\ket{s}$. 

\subsection{Vertex excitations}
\label{sec:vExcitations}

The eigenvalue 0 for the vertex operator in (\ref{vo}) corresponds to the eigenvalue 1 for one of the orthogonal vertex operators, $A_v^{[\alpha]}$ in (\ref{voperps}) with $\alpha\neq 0$ mod $n$. 
First we look at the possibility for a single vertex operator to be excited, or an isolated vertex excitation, at the vertex $v'$ by computing 
\be
\Tr_\mathcal{H}\left(A_{v'}^{\left[\alpha\right]}\prod\limits_{v\in V-\{v'\}} A_v\prod\limits_{e \in E}B_e \right) = 
\frac{(\mbox{GSD})}{q}\sum_{b=0}^{q-1}\omega_q^{\alpha b},
\label{TrvExcitations}
\ee
which does not vanish only when $\alpha=0$ mod $q$. 
It implies that $k-1$ isolated vertex excitations exist corresponding to $\alpha\in\{q, 2q, \cdots, (k-1)q\}$. 
As we will see later, the remaining possibilities $\alpha\neq 0$ mod $q$ contribute to deconfined excitations.  

\paragraph{Excited states $z_v^\beta\ket{\mbox{GS}\,s}$:}
As in the previous subsection, to create the isolated vertex excitations let us pick the state $z_v^\beta\ket{\mbox{GS}, s}$ with $\beta\in\{1,\cdots,m-1\}$. 
Since $z_v^k$ generates a local symmetry, the cases $\beta\in\{1,\cdots,k-1\}$ are candidates for the independent excitations. 
%
The relation
\be
z_v^\beta A_v^{(j)}=\omega_m^{\beta p j}A_v^{(j)}z_v^\beta
\ee
together with $\omega_m^{\beta pj}=\omega_k^{\beta j}=\omega_n^{\beta q j}$ leads to 
\be
z_v^\beta A_v=A_v^{[\beta q]}z_v^\beta.
\ee
Then, we have 
\be
z_v^\beta\ket{\mbox{GS}\,s}=\omega_m^{\beta a_v}\sqrt{\cN}\,A_v^{[\beta q]}\left(\prod_{v'\in V-\{v\}}A_{v'}\right)\ket{s}. 
\ee
Note that $A_v^{[\beta q]}$ is orthogonal to $A_v$ only when $\beta\neq 0$ mod $k$. 
Thus, $z_v^\beta\ket{\mbox{GS}\,s}$ with $\beta\in\{1,2,\cdots, k-1\}$ are independent isolated vertex excitations with the energy $E_0+1$ (first excited states).


\paragraph{Excited states $Z_e^\gamma\ket{\mbox{GS}\,s}$:}
Next we turn our attention to the deconfined vertex excitations that occur in pairs. 
These are similar to the abelian quantum double models, and hence we first pick the states $Z_e^\gamma\ket{\mbox{GS}\, s}$ for $\gamma\in\{1, 2,\cdots,n-1\}$.
Since $Z_e^{\tilde{n}}=Z_e^{\kti q}$ generates a local symmetry as mentioned around (\ref{ntilde}), we may consider the cases $\gamma\in\{1,2,\cdots, \kti q-1\}$. 
Furthermore, 
\be\label{qmult}
Z^{qj}_e = z_{v_1}^{-pj}\left[z_{v_1}^{pj}Z_e^{qj}z_{v_2}^{-pj}\right]z_{v_2}^{pj}=z_{v_1}^{-pj}z_{v_2}^{pj}B_e^{(j)} \quad \mbox{for $j\in\{1, \cdots,k-1\}$}
\ee
implies that whenever $\gamma$ is a multiple of $q$, $Z_e^\gamma\ket{\mbox{GS}\,s}$ reduces to a composition of two isolated vertex excitations on $v_1$ and $v_2$, 
since $B_e^{(j)}$ can be written as a linear combination of the edge operator $B_e$ and its orthogonal complements $B_e^{[\alpha]}$ in (\ref{orthoBe}). 
Thus, we find candidates for independent deconfined vertex excitations as those for $\gamma\in \{1,2,\cdots, q-1\}$. 


From (\ref{xz_XZ}) we obtain for $e\in L_v^\pm$
\be
Z_e^\gamma A^{(j)}_v = \omega_n^{\pm\gamma pj} A_v^{(j)}Z_e^\gamma \qquad (j\in \mathbb{Z}_n),
\ee
and thus
\be
Z_e^\gamma A_v = A_v^{[\pm \gamma p]}Z_e^\gamma.
\ee
Here, $\gcd(p,q)=\gcd(\xi\pti,q)=1$ provides $\gcd(\pti,q)=1$. This and $\gcd(\kti,\pti)=1$ give $\gcd(\kti q, \pti)=1$, which leads to
$\omega_n^{\gamma p} =\omega_{\kti q}^{\gamma \pti}\neq 1$ and thus $A_v^{[\pm\gamma p]}$ are orthogonal to $A_v$ for any $\gamma$ in the above range.
For an edge $e$ belonging to $L_{v_1}^-$ and $L_{v_2}^+$ as in Fig.~\ref{edge}, we explicitly see 
\be
Z_e^\gamma\ket{\mbox{GS}\,s}=\omega_q^{p\gamma\sum_{L=1}^{B_1}\delta_{e,\hat{e}_L}b_L}\sqrt{\cN}\,A_{v_1}^{[-\gamma p]}A_{v_2}^{[\gamma p]}\left(\prod_{v'\in V-\{v_1,v_2\}}A_{v'}\right)\ket{s},
\label{ZegammaGSs}
\ee
which indicates that $Z_e^\gamma\ket{\mbox{GS}\,s}$ with $\gamma\in \{1,2,\cdots,q-1\}$ are second excited states with the energy $E_0+2$. 
The excitations are paired and occur at the both ends of the edge $e$, namely the vertices $v_1$ and $v_2$. 
Now it is clear that among the possibilities of $\alpha\neq  0$ mod $q$ mentioned below (\ref{TrvExcitations}), $q-1$ of them are independent and corresponds to the excitations (\ref{ZegammaGSs}).

\begin{figure}[H]
\centering
\captionsetup{width=.8\linewidth}
\begin{tikzpicture}
\fill (0,0) circle [radius=2pt];
\fill (-2,-1) circle [radius=2pt];
\fill (2,0) circle [radius=2pt];
\draw [->,thick] (-2,-1)--(-1,-0.5);
\draw[ -,thick] (-1,-0.5)--(0,0);
\draw [->,thick] (0,0)--(1,0);
\draw [-,thick] (1,0)--(2,0);
\node (v1) at (-2.3,-1.3) {$v_1$};
\node (v2) at (2.3,0.3) {$v_2$};
\node (v3) at (0,0.5) {$v_3$};
\node (e1) at (-1,-1) {$e_1$};
\node (e2) at (1,-0.5) {$e_2$};
\draw [-,thick] (-2,-1)--(-2.3,-0.7);
\draw [-,thick] (-2,-1)--(-2,-1.5);
\draw [-,thick] (0,0)--(-0.3,0.3);
\draw [-,thick] (0,0)--(0,-0.4);
\draw [-,thick] (2,0)--(1.7,-0.3);
\draw [-,thick] (2,0)--(2.4,-0.1);
\end{tikzpicture}
\caption{Three vertices $v_1$, $v_2$ and $v_3$ are connected by two edges $e_1$ and $e_2$ such that $e_1$ is directed from $v_1$ to $v_3$, and $e_2$ is directed from $v_3$ to $v_2$.}
\label{fig:e12v123}       
\end{figure}
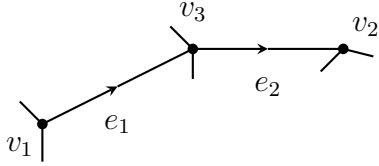
Likewise, for two edges ($e_1$ and $e_2$) and three vertices ($v_1$, $v_2$ and $v_3$), where $e_1\in L_{v_1}^-, L_{v_3}^+$ and $e_2\in L_{v_3}^-, L_{v_2}^+$ as in Fig.~\ref{fig:e12v123}, 
consecutive two excitations read 
\bea
Z_{e_1}^{\gamma_1}Z_{e_2}^{\gamma_2}\ket{\mbox{GS}\,s} & = &  \omega_q^{p\sum_{a=1}^2\gamma_a\sum_{L=1}^{B_1}\delta_{e_a,\hat{e}_L}b_L}\sqrt{\cN} \nn \\
& & \times A_{v_1}^{[-\gamma_1p]}A_{v_3}^{[(\gamma_1-\gamma_2)p]}A_{v_2}^{[\gamma_2p]}\left(\prod_{v'\in V-\{v_1,v_2,v_3\}}A_{v'}\right)\ket{s}
\label{Ze1Ze2}
\eea
with $\gamma_1, \gamma_2\in \{1,2,\cdots,q-1\}$. Note that the relation 
\be
Z_{e_1}^{\gamma_1}A_{v_3}^{[-\gamma_2 p]}=A_{v_3}^{[(\gamma_1-\gamma_2)p]}Z_{e_1}^{\gamma_1}
\ee 
holds.
When $\gamma_1=\gamma_2$, the excitation at $v_3$ disappears and (\ref{Ze1Ze2}) become second excited states.  
Based on this observation, we can claim a general statement. 
Let $P$ be an arbitrary path directed from the vertex $v_1$ to the vertex $v_2$ on the graph. 
For the Wilson line operator along $P$,\footnote{$(e|P)$ is a sign factor defined similarly to $(e|C)$ at (\ref{ZC}).} 
\be
Z(P)\equiv \prod_{e\in P}Z_e^{(e|P)},
\label{ZP}
\ee 
the states $Z(P)^\gamma\ket{\mbox{GS}\,s}$ are at the second excited level with the energy $E_0+2$ for $\gamma\in \{1,2,\cdots,q-1\}$. 
The excitations occur at the endpoints of $P$, $v_1$ and $v_2$.


\subsection{Case of $m=4, n=6$}
\label{sec:case}
For illustrative purposes of the peculiarities of the excited states, we concretely present the case of $m=4$ and $n=6$, i.e., $k=p=2$ and $q=3$. 
In this case there are six mutually orthogonal vertex operators, $A_v^{[\alpha]}$, $\alpha\in\{0,1,\cdots, 5\}$, and two mutually orthogonal edge operators, $B_e^{[\alpha]}$, $\alpha\in\{0,1\}$. These operators are given in (\ref{voperps}) and (\ref{orthoBe}) respectively and for clarity we write their full expressions here.
\bea
A_v^{[0]} & = & \frac{1}{6}\left[1_v1_{L_v}+x^2_v\bX^2_{L_v} + 1_v\bX ^4_{L_v} + x^2_v1_{L_v} + 1_v\bX ^2_{L_v} + x^2_v \bX ^4_{L_v}\right],\nonumber \\
A_v^{[1]} & = & \frac{1}{6}\left[1_v1_{L_v}+\omega_6x^2_v \bX ^2_{L_v} + \omega_6^2 1_v\bX ^4_{L_v} + \omega_6^3 x^2_v 1_{L_v}+ \omega_6^4 \bX ^2_{L_v} + \omega_6^5 x^2_v \bX ^4_{L_v}\right],\nonumber \\ 
A_v^{[2]} & = & \frac{1}{6}\left[1_v1_{L_v}+\omega_3x^2_v \bX ^2_{L_v} + \omega_3^2 1_v\bX ^4_{L_v} + x^2_v1_{L_v} + \omega_3 \bX ^2_{L_v} + \omega_3^2 x^2_v \bX ^4_{L_v}\right],\nonumber \\
A_v^{[3]} & = & \frac{1}{6}\left[1_v1_{L_v}-x^2_v \bX ^2_{L_v} + 1_v\bX ^4_{L_v} - x^2_v1_{L_v} + 1_v\bX ^2_{L_v} - x^2_v \bX ^4_{L_v}\right],\nonumber \\ 
A_v^{[4]} & = & \frac{1}{6}\left[1_v1_{L_v}+\omega_3^2x^2_v \bX ^2_{L_v} + \omega_3 1_v\bX ^4_{L_v} + x^2_v1_{L_v} + \omega_3^2 \bX ^2_{L_v} + \omega_3 x^2_v \bX ^4_{L_v}\right],\nonumber \\
A_v^{[5]} & = & \frac{1}{6}\left[1_v1_{L_v}+\omega_6^5x^2_v \bX ^2_{L_v} + \omega_6^4 1_v\bX ^4_{L_v} + \omega_6^3 x^2_v1_{L_v} + \omega_6^2 \bX ^2_{L_v} + \omega_6 x^2_v \bX ^4_{L_v}\right],\nonumber 
\eea
and 
\bea
B_e^{[0]} & = & \frac{1}{2}\left[1_{v_1}1_e1_{v_2} + z_{v_1}^2Z_e^3z_{v_2}^2\right], \nonumber \\
B_e^{[1]} & = & \frac{1}{2}\left[1_{v_1}1_e1_{v_2}  - z_{v_1}^2Z_e^3z_{v_2}^2\right]. \nonumber
\eea
Here, $\omega_6=e^{\frac{2\pi\mathrm{i}}{6}}$, $\omega_3=e^{\frac{2\pi\mathrm{i}}{3}}$, $x_v^4=z_v^4=1_v$ and $X_e^6=Z_e^6=1_e$. 
While the ground states are the +1 eigenstates of $A_v^{[0]}$ and $B_e^{[0]}$, the excited states are +1 eigenstates of the remaining orthogonal operators, as discussed in sections~\ref{sec:eExcitations} and \ref{sec:vExcitations}. We will exhaust them using the results of sections~\ref{sec:eExcitations} and \ref{sec:vExcitations}. 

The first excited states are given by $X_e\ket{\mbox{GS}\,s}$ and $z_v\ket{\mbox{GS}\,s}$ for any $e\in E$ and $v\in V$, 
which are isolated (immobile) and appear when $k\neq 1$ in general. 
They are the isolated edge and vertex excitations and are easily seen as the +1 eigenstates of the operators, $B_e^{[1]}$ and $A_v^{[3]}$, respectively.
%
The second excited states are given by $Z(P)\ket{\mbox{GS}\,s}$ and $Z(P)^2\ket{\mbox{GS}\,s}$ with $Z(P)$ being (\ref{ZP}) for any open path on the graph $G$, which are deconfined (mobile) 
and appear when $q\neq 1$ in general. 

The operators $Z(P)$ and $Z(P)^2$ create vertex excitations at the endpoints of the path $P$, which are the +1 eigenstates of the vertex operators, $A_v^{[2]}$ and $A_v^{[4]}$. Finally appending the operators $z_v$ at the end points of the path operators, $Z(P)$ and $Z(P)^2$ we create the +1 eigenstates of the operators, $A_v^{[1]}$ and $A_v^{[5]}$. 
Interestingly, these combined excitations are also at the second excited level.\footnote{
We did not considered such combined excitations in section~\ref{sec:vExcitations} because the analysis seems to be complicated for general $m$ and $n$.}   

Note that there are no deconfined excited states created by $X_e$.
This realizes the exchange phases of `anyons' which are not exactly the same as what appear in the toric code or the two-dimensional quantum double models 
of Kitaev (see the following subsections).

\subsection{Anyon-like excitations and topological EE}
\label{sec:stat}

In this section we discuss anyon-like excitations and their relevance to the topological EE computed in sections~\ref{sec:EEGSs} and \ref{sec:EEGSab}. 

Let us first recall the toric code model~\cite{kitaev}. 
The toric code model is defined on the square lattice\footnote{The toric code models are well-defined on any triangulation of the two dimensional space, and the square lattice is usually chosen for simplicity.} 
with the Pauli spin operators, $\bar{X}_e$ and $\bar{Z}_e$, acting on each link.\footnote{
We put the bar to the operators of the toric code model in order to distinguish the operators in our model.}  
The Hamiltonian consists of two kinds of interaction terms -- the star term $A_v$ consisting of $\bar{X}_e$'s and the plaquette term $B_p$ consisting of $\bar{Z}_e$'s. 
The former energetically imposes the Gauss law constraints, and the latter gives a standard gauge kinetic term on the lattice. 
It is analogous to the $\mathbb{Z}_2$ lattice gauge theory.  
There are two kinds of deconfined excitations. One is `electric excitations' which are constructed by Wilson line operators of $\bar{Z}_e$ acting on the ground states.
Associated to a path on the (original) lattice, the corresponding Wilson line operator is defined by the product of $\bar{Z}_e$'s along the path. 
The excitations occur at the endpoints of the path, which can be interpreted as electric charges. 
The other is `magnetic excitations' constructed by acting 't Hooft line operators on the ground states. 
Associated to a path on the dual lattice, the corresponding 't Hooft line operator is defined by the product of $\bar{X}_e$'s on the edges $e$ intersecting with the path. 
The excitations appear at the endpoints of the path, which can be interpreted as magnetic fluxes. 
When an electric charge moves around a magnetic flux (and vice versa), an anyon phase appears due to the Aharonov-Bohm effect. 
See Fig.~\ref{fig:thooft} for an example of `t Hooft and Wilson line operators.  
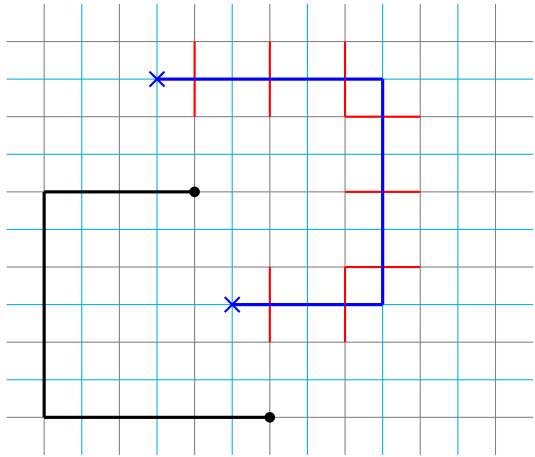
\begin{figure}[h]
\centering
\captionsetup{width=.8\linewidth}
\begin{tikzpicture}
\foreach \x in {-3,-2,-1,0,1,2,3}
   \draw[-,thin, color=gray] (\x,2.5)--(\x,-3.5);
\foreach \x in {-3,-2,-1,0,1,2}
   \draw[-,thin, color=gray] (3.5,\x)--(-3.5,\x);    
\foreach \x in {-2.5,-1.5,-0.5,0.5,1.5,2.5}
   \draw[-,thin, color=cyan] (\x,2.5)--(\x,-3.5);
\foreach \x in {-2.5,-1.5,-0.5,0.5,1.5}   
   \draw[-,thin, color=cyan] (3.5,\x)--(-3.5,\x);
\draw [-,thick, color=red] (-1,2)--(-1,1);
\draw [-,thick, color=red] (0,2)--(0,1);
\draw [-,thick, color=red] (1,2)--(1,1);
\draw [-,thick, color=red] (1,1)--(2,1);
\draw [-,thick, color=red] (1,0)--(2,0);
\draw [-,thick, color=red] (1,-1)--(2,-1);
\draw [-,thick, color=red] (1,-1)--(1,-2);
\draw [-,thick, color=red] (0,-1)--(0,-2);
\draw [-,very thick, color=blue] (-1.5,1.5)--(1.5,1.5);
\draw [-,very thick, color=blue] (1.5,1.5)--(1.5,-1.5);
\draw [-,very thick, color=blue] (1.5,-1.5)--(-0.5,-1.5); 
\draw [-,thick, color=blue] (-1.6,1.6)--(-1.4,1.4);
\draw [-, thick, color=blue] (-1.6,1.4)--(-1.4,1.6);
\draw [-, thick, color=blue] (-0.6,-1.4)--(-0.4,-1.6);
\draw [-, thick, color=blue] (-0.6,-1.6)--(-0.4,-1.4);
\draw[-,very thick] (-1,0)--(-3,0);
\draw[-,very thick] (-3,0)--(-3,-3);
\draw[-,very thick] (-3,-3)--(0,-3);
\fill (-1,0) circle [radius=2pt];
\fill (0,-3) circle [radius=2pt];
\end{tikzpicture}
\caption{An example of 't Hooft and Wilson line operators in the toric code model on the square lattice. 
For the original lattice drawn in the gray lines, the dual lattice is drawn in the light blue lines.   
The blue line represents a path on the dual lattice, and its associated 't Hooft line operator is  
given by the product of $\bar{X}_e$'s on the red edges. 
The blue crosses represent the ends of the 't Hooft line, at which magnetic excitations occur. 
The black line is a path on the original lattice. The product of $\bar{Z}_e$'s along the path gives the associated Wilson line. 
Electric excitations appear at the endpoints of the path (the black dots).}
\label{fig:thooft}      
\end{figure}

Clearly the electric excitations correspond to the Wilson line operators $Z(P)^\gamma$ ($\gamma=1,2,\cdots,q-1$) acting on the ground states in our case. 
However, there seems to be no counterpart to the magnetic excitations in excitations discussed in sections~\ref{sec:eExcitations} and \ref{sec:vExcitations}. 
We see that the $\mathbb{Z}_q$ magnetic fluxes $pb_L$'s on the ground states $\ket{\mbox{GS}\,s}$ play an analogous role to the magnetic excitations, 
except the point that the magnetic fluxes do not cost any energy, or they are condensed into the ground state.
In our case, since the plaquette terms of $Z_e$'s are absent in the Hamiltonian (\ref{H}), the ground states can accommodate the 
zero-energy magnetic fluxes.\footnote{Interestingly, adding the plaquette terms of $Z_e^q$'s rather than $Z_e$'s to the Hamiltonian does not alter the ground states. 
} 
From (\ref{ZCL_GSs}), after one of the endpoints of the Wilson line circulates along the closed path $C_L$, it acquires the Aharonov-Bohm phase\footnote{
A similar phenomenon is observed in {\it topological flux phases} in the {\it string net models}~\cite{levin_wen}. 
Although the string net models normally allow ground states with zero flux, the topological flux phases are realized 
by modifying the Hamiltonians so that nonzero flux states are energetically favored~\cite{patel}. 
} 
\be
\omega_q^{p\gamma b_L\,(\hat{e}_L|C_L)}. 
\label{ABphase}
\ee
Since a general closed path on the graph is a linear combination of $C_L$'s with the coefficients $\pm 1$, the phase appearing after moving along the general path is given by the product of 
the phases for each $C_L$. 
In the phase (\ref{ABphase}), $\gamma$ and $b_L$ are $\mathbb{Z}_q$-valued (including the trivial case), which leads to the total quantum dimension 
$D=\sqrt{q^2}=q$. This accounts for the topological EE term obtained in section~\ref{sec:EEGSs}. 

On the other hand, the ground states $\ket{\mbox{GS}[\alpha,\beta]}$ have the $\mathbb{Z}_q$ electric flux $\beta_L(\hat{e}_L|C_L)$ along $C_L$ as shown in section~\ref{sec:GSab}. 
$\beta_L$ corresponds to $\gamma$ in the above. 
Thus the same process acquiring the anyon phase (\ref{ABphase}) occurs by inserting the {\it local operator} $X_{\hat{e}_L}^{-pbk\,(\hat{e}_L|C_L)}$ ($b\in \mathbb{Z}_q$). 
As is seen from (\ref{Xe_GSab}),
\be
X_{\hat{e}_L}^{-pb_Lk\,(\hat{e}_L|C_L)}  \ket{\mbox{GS}[\alpha,\beta]}=\omega_q^{pb_L\beta_L (\hat{e}_L|C_L)} \ket{\mbox{GS}[\alpha,\beta]}. 
\ee
This operator is local and does not contribute to the topological EE, which explains the reason why the topological EE vanishes for $\ket{\mbox{GS}[\alpha,\beta]}$ in the result (\ref{EEabf}). 


\subsection{Exchange statistics on graphs}
\label{sec:otherstat}

Before closing this section, we mention about connections to the analysis in \cite{agraph1,agraph11,agraph12} to identify exchange statistics that are peculiar to graphs. 
There, quantum particles sit on vertices of graphs, and it is investigated which phases can appear for various patterns of the exchange of the particles. 
The setting is different from the case of the abelian toric code models, 
in which 
an anyon phase appears only when an electric excitation at a vertex of the (original) lattice moves around a magnetic one at a vertex of the {\it dual lattice} and vice versa.      
When an electric excitation moves around another electric one, no phase appears. 

Similarly, in our case, movable excitations at vertices are only of the electric type, and their exchange as in Fig.~\ref{fig:exchange_loop} does not provide a nontrivial phase. 
However, in case that the ground state $\ket{\mbox{GS}\,s}$ has magnetic flux penetrating the region enclosed by the loop $C: v_1\to v_2\to v_3\to v_4\to v_1$, 
the Aharonov-Bohm phase appears as already mentioned. 
%
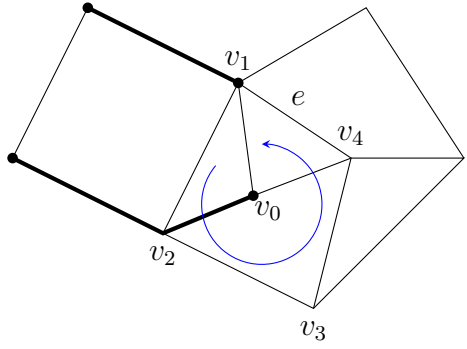
\begin{figure}[h]
\centering
\captionsetup{width=.8\linewidth}
\begin{tikzpicture}
\fill (0,0) circle [radius=2pt];
\fill (1,2) circle [radius=2pt];
\fill (3,1) circle [radius=2pt];
\fill (3.2,-0.5) circle [radius=2pt];
\draw [-,ultra thick] (1,2)--(3,1);
\draw [-,ultra thick] (0,0)--(2,-1)--(3.2,-0.5);
\draw [-,thin] (1,2)--(0,0);
\draw [-,thin] (2,-1)--(3,1)--(3.2,-0.5)--(4.5,0)--(6,0)--(4,-2)--(2,-1);
\draw [-,thin] (4,-2)--(4.5,0)--(3,1)--(4.7,2)--(6,0); 
\node (v1) at (3,1.3) {$v_1$};
\node (v2) at (2,-1.3) {$v_2$};
\node (v3) at (4,-2.3) {$v_3$};
\node (v4) at (4.5,0.3) {$v_4$};
\node (v0) at (3.4,-0.7) {$v_0$};
\node (e) at (3.8,0.8) {$e$};
\draw [->,blue] (2.7,-0.1) arc (-220:90:0.8cm);
\end{tikzpicture}
\caption{Two Wilson lines on a graph. The thick lines denote the Wilson lines. The orientations are suppressed. When the excitation at the vertex $v_1$ move around the excitation at $v_0$ 
along the loop $C: v_1\to v_2\to  v_3\to v_4\to v_1$ (as the blue arrow indicates), there appears no anyon phase because they are both electric excitations. 
Interestingly, if there is an isolated excitation on some edge $e$ on the loop $C$, the electric excitation moves along $C$ on $X_e^\beta\ket{\mbox{GS}\,s}$ ($\beta=\{1,\cdots,k-1\}$) provides a phase.}
\label{fig:exchange_loop}       
\end{figure}
Interestingly, even when the ground state does not have the magnetic flux, and instead an isolated excitation $X_e^\beta$ ($\beta\in \{1,\cdots, k-1\}$) exists on some edge $e$ on the loop, 
a nontrivial phase appears as 
\be
Z(C)^\gamma X_e^\beta \ket{\mbox{GS}\,s}=\omega_n^{\beta\gamma \,(e|C)}X_e^\beta Z(C)^\gamma\ket{\mbox{GS}\,s}=  \omega_n^{\beta\gamma \,(e|C)}X_e^\beta \ket{\mbox{GS}\,s},
\ee
where $\gamma\in\{1,\cdots,q-1\}$, and the last equality comes from (\ref{ZCL_GSs}) with $b_L=0$ (no magnetic flux). 
This can be thought of as a generalization of the Aharnov-Bohm effect in the presence of the 0-holonomy operator or the edge operator. 

In addition, the exchange on a T- (or Y-) junction is considered in \cite{agraph1,agraph11,agraph12}, where 
two particles at the vertices $(v_3,v_2)$ move as $(v_3,v_2)\to (v_3,v_4)\to (v_2,v_4)\to (v_1,v_4)\to (v_1,v_2) \to (v_1,v_3) \to (v_2,v_3)$ in Figs.~\ref{fig:exchange_T} and \ref{fig:exchangeprocess}. 
In our case no phase appears in the process as seen below. 
%
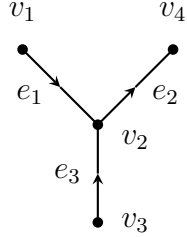
\begin{figure}[h]
\centering
\captionsetup{width=.8\linewidth}
\begin{tikzpicture}
\fill (0,0) circle [radius=2pt];
\fill (-1,1) circle [radius=2pt];
\fill (1,1) circle [radius=2pt];
\fill (0,-1.3) circle [radius=2pt];
\draw [->,thick] (-1,1)--(-0.5,0.5);
\draw [-,thick] (-0.5,0.5)--(0,0);

\draw[->,thick] (0,-1.3)--(0,-0.65);
\draw[-,thick] (0,-0.65)--(0,0);

\draw [->,thick] (0,0)--(0.5,0.5);
\draw [-,thick] (0.5,0.5)--(1,1);
\node (v1) at (0.5,-1.3) {$v_3$};
\node (v2) at (0.5,-0.2) {$v_2$};
\node (v3) at (-1,1.5) {$v_1$};
\node (v4) at (1,1.5) {$v_4$};
\node (e1) at (-0.9,0.4) {$e_1$};
\node (e2) at (0.9,0.4) {$e_2$};
\node (e3) at (-0.4,-0.65) {$e_3$};
\end{tikzpicture}
\caption{A graph of a T- (or Y-) junction. 
}
\label{fig:exchange_T}       
\end{figure}

\begin{figure}[h]
\centering
\captionsetup{width=.8\linewidth}
\begin{tikzpicture}
\fill[red] (0,-2) circle [radius=3pt];
\fill[blue] (0,-3.3) circle [radius=3pt];
\draw [-,thick] (-1,-1)--(0,-2);

\draw[-,thick] (0,-3.3)--(0,-2);

\draw [-,thick] (0,-2)--(1,-1);
%

\draw [->,thick] (2,-2/3)--(3,-1/3);

\fill[red] (5,1) circle [radius=3pt];
\fill[blue] (4,-1.3) circle [radius=3pt];
\draw [-,thick] (3,1)--(4,0);

\draw[-,thick] (4,-1.3)--(4,0);

\draw [-,thick] (4,0)--(5,1);
%

\draw [->,thick] (5.75,0)--(6.75,0);

\fill[blue] (8,0) circle [radius=3pt];
\fill[red] (9,1) circle [radius=3pt];
%
\draw [-,thick] (7,1)--(8,0);

\draw[-,thick] (8,-1.3)--(8,0);

\draw [-,thick] (8,0)--(9,1);
%

\draw [->,thick] (9.75,0)--(10.75,0);

\fill[blue] (11,1) circle [radius=3pt];
\fill[red] (13,1) circle [radius=3pt];
%
\draw [-,thick] (11,1)--(12,0);

\draw[-,thick] (12,-1.3)--(12,0);

\draw [-,thick] (12,0)--(13,1);
%

\draw [->,thick] (12,-2)--(12,-3);

\fill[red] (12,-4) circle [radius=3pt];
\fill[blue] (11,-3) circle [radius=3pt];
%
\draw [-,thick] (11,-3)--(12,-4);

\draw[-,thick] (12,-5.3)--(12,-4);

\draw [-,thick] (12,-4)--(13,-3);
%

\draw [->,thick] (10.75,-4)--(9.75,-4);

\fill[blue] (7,-3) circle [radius=3pt];
\fill[red] (8,-5.3) circle [radius=3pt];
\draw [-,thick] (7,-3)--(8,-4);

\draw[-,thick] (8,-5.3)--(8,-4);

\draw [-,thick] (8,-4)--(9,-3);
%

\draw [->,thick] (6.75,-4)--(5.75,-4);

\fill[blue] (4,-4) circle [radius=3pt];
\fill[red] (4,-5.3) circle [radius=3pt];
\draw [-,thick] (3,-3)--(4,-4);

\draw[-,thick] (4,-5.3)--(4,-4);

\draw [-,thick] (4,-4)--(5,-3);
%
\end{tikzpicture}
\caption{The exchange process of two particles (the red and blue dots) in a T-(Y-) junction of the graph. For simplicity the orientations and other vertices are suppressed. 
}
\label{fig:exchangeprocess}       
\end{figure}
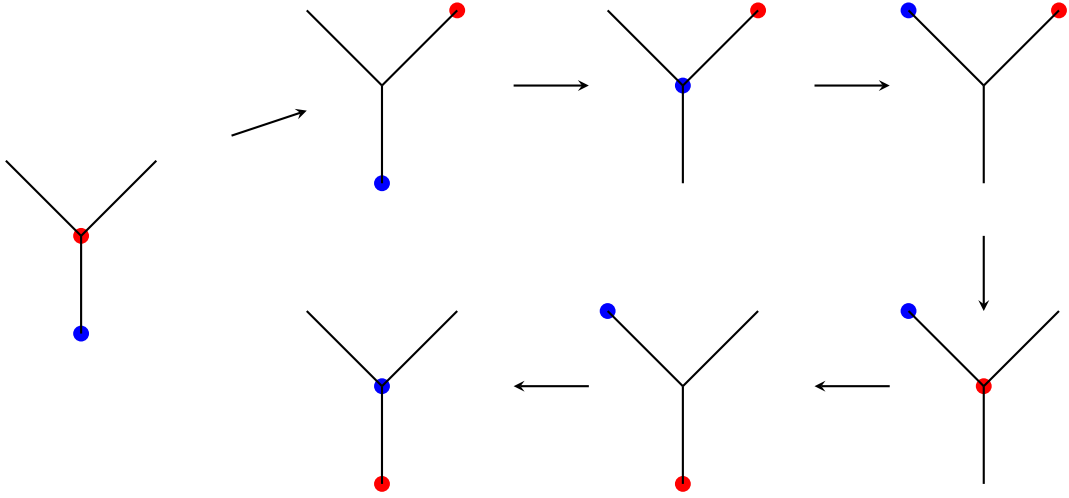

Let us consider two deconfined excitations of strings $Z(P_1)^{\gamma_1}$ and $Z(P_2)^{\gamma_2}$ ($\gamma_1, \gamma_2\in\{1,\cdots, q-1\}$) on any fixed state $\ket{\psi}$. 
Suppose the T-(or Y-) junction is a part of the graph and the red and blue dots in Fig.~\ref{fig:exchangeprocess} are endpoints of $Z(P_1)^{\gamma_1}$ and $Z(P_2)^{\gamma_2}$, respectively. 
When $e_3\in P_1$,\footnote{We can similarly show for other configurations of the path.} the initial state is written as 
\be
\ket{\psi_i}=Z_{e_3}^{\gamma_1}\cdots\ket{\psi},
\label{psi_i}
\ee
where $(\cdots)$ expresses the strings outside the T-(Y-) junction and does not change in the process (see Fig.~\ref{fig:initial_psi}). 
%
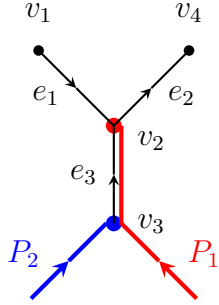
\begin{figure}[h]
\centering
\captionsetup{width=.8\linewidth}
\begin{tikzpicture}
\fill[red] (0,0) circle [radius=3pt];
\fill (-1,1) circle [radius=2pt];
\fill (1,1) circle [radius=2pt];
\fill[blue] (0,-1.3) circle [radius=3pt];
\draw [->,thick] (-1,1)--(-0.5,0.5);
\draw [-,thick] (-0.5,0.5)--(0,0);

\draw[->,thick] (0,-1.3)--(0,-0.65);
\draw[-,thick] (0,-0.65)--(0,0);

\draw [->,thick] (0,0)--(0.5,0.5);
\draw [-,thick] (0.5,0.5)--(1,1);
\node (v1) at (0.5,-1.3) {$v_3$};
\node (v2) at (0.5,-0.2) {$v_2$};
\node (v3) at (-1,1.5) {$v_1$};
\node (v4) at (1,1.5) {$v_4$};
\node (e1) at (-0.9,0.4) {$e_1$};
\node (e2) at (0.9,0.4) {$e_2$};
\node (e3) at (-0.4,-0.65) {$e_3$};
\draw [->, red, ultra thick] (1.1,-2.3)--(0.6,-1.8);
\draw [-, red, ultra thick] (0.6,-1.8)--(0.1,-1.3)--(0.1,0);
\draw [->, blue, ultra thick] (-1.1,-2.3)--(-0.6,-1.8);
\draw [-,blue, ultra thick] (-0.6,-1.8)--(-0.1,-1.3);
\node (g1) at (1.2,-1.7) {\textcolor{red}{$P_1$}};
\node (g2) at (-1.2,-1.7) {\textcolor{blue}{$P_2$}};
\end{tikzpicture}
\caption{An initial configuration of the strings $Z(P_1)^{\gamma_1}$ (red line) and $Z(P_2)^{\gamma_2}$ (blue line). 
}
\label{fig:initial_psi}       
\end{figure}
%
We trace the path of each of the strings in the exchange process. The endpoint of the string $Z(P_1)^{\gamma_1}$ (the red dot) moves as $v_2\to v_4\to v_2 \to v_3$ which gives the additional contribution to the string 
\be
Z_{e_3}^{-\gamma_1}Z_{e_2}^{-\gamma_1}Z_{e_2}^{\gamma_1}=Z_{e_3}^{-\gamma_1},
\ee
whereas the endpoint of the string $Z(P_2)^{\gamma_2}$ (the blue dot) moves as $v_3\to v_2\to v_1\to v_2$ which yields
\be
Z_{e_1}^{\gamma_2} Z_{e_1}^{-\gamma_2} Z_{e_3}^{\gamma_2}=Z_{e_3}^{\gamma_2}.
\ee
Thus, $Z_{e_3}^{-\gamma_1+\gamma_2}$ is obtained in total. Including this, the final state becomes
\be
\ket{\psi_f}=Z_{e_3}^{-\gamma_1+\gamma_2}\ket{\psi_i}=Z_{e_3}^{\gamma_2}\cdots\ket{\psi}.
\label{psi_f}
\ee
(\ref{psi_f}) is the same as (\ref{psi_i}) after $\gamma_2$ is replaced by $\gamma_1$, which realizes the exchange without a phase. 
Note that this holds for any state $\ket{\psi}$\footnote{As a remark we would like to add that it is possible to obtain exchange phases in this process if we move the deconfined vertex excitations with $Z_e^{q-\gamma}$ instead of $Z_e^{-\gamma}$. This is allowed as $\gamma\in\{1,2,\cdots, q-1\}$. However this exchange process is not adiabatic as $Z_e^q$ (See (\ref{qmult})) creates extra vertex excitations changing the energy of the state during the exchange process. The phase, upon exchanging the deconfined vertex excitations in the background of three isolated edge excitations created by $X_{e_1}^{\beta_1}$, $X_{e_2}^{\beta_2}$ and $X_{e_3}^{\beta_3}$ is,  $\omega_n^{q\beta_1+q \beta_2 +2q\beta_3}$.}. 

\section{Discussion}
\label{sec:Disc}
\subsection{Summary}
In this paper we have initiated a detailed study of abelian gauge theories on graphs that host
quantum phases not classified as phases of spontaneous symmetry breaking with local order parameters.   
In general the models realize such quantum phases with a mixture of topological and 
extensive aspects.\footnote{According to Theorem IV.6 in \cite{pt1}, the GSD is given by
\[
\mbox{GSD} = \prod_n \left|H^n(C,H_n(G)\right| 
 = \underbrace{\left|\mbox{Hom}(H_0(C),H_0(G)\right|}_{p^{|V|}} \underbrace{\left|\mbox{Hom}(H_1(C),H_1(G)\right|}_{p^{|E|-|V|+1}} 
\]
since the Ext and Tor functors are null.  
This implies that the extensive part of the GSD, $p^{|V|}$, can also be regarded as a topological contribution from the mathematical perspective.
However, we should remark that each of the ground states contributing to the extensive part is fragile under local perturbations, 
as mentioned in the last paragraph of this section, defying the physicist's definition of topological phases. 
} 
%
They possess features that are reminiscent of the two-dimensional quantum double models of Kitaev \cite{kitaev}, 
with some subtle differences in their properties, notably in the GSD and in the nature of the anyonic excitations. 

In some cases occurring for specific families of $m$ and $n$ values, our result reads:
\begin{itemize}
\item 
When $p=1$, $n$ is an integer multiple of $m$: $n=mq$. 
We obtain the purely topological case in which the GSD reduces to $q^{B_1}$. The EE for the ground states, $\ket{\mbox{GS}\,s}$, includes the constant term (topological EE) as $-\gamma=-\log_2q$, accounted for by the anyons in the model. 
%
\item
When $q=1$, $m$ is an integer multiple of $n$: $m=np$. 
The GSD is extensive as seen by the expression $p^{|V|}$. 
For the ground states $\ket{\mbox{GS}\,s}$, the topological EE vanishes. 
\item 
When $m=n$, there is a unique ground state obeying the area law for the EE. 
It is also interesting to note that the Hamiltonian for $m=n=2$ is unitarily equivalent to the {\it cluster state Hamiltonians} representing a $\mathbb{Z}_2\times\mathbb{Z}_2$ {\it symmetry protected topological} (SPT) phase \cite{clusterM}, 
when the graph forms either an open or closed chain.
\item
When $k=1$, $m=p$ and $n=q$ are co-prime to each other. 
We do not have valid homomorphisms, $\partial^{\left[l\right]}$ between $\mathbb{Z}_n$ and $\mathbb{Z}_m$ except for the trivial homomorphism. 
%
In this case, the edge operators become trivial ($B_e=1$), and the vertex operators reduce to $A_v=\frac{1}{q}\sum_{j=0}^{q-1}1_v\bX_{L_v}^{pj}$ which 
impose the Gauss law constraints of the pure gauge theory. 
Although only the operators $X_e$ appear in the Hamiltionian, the result for the GSD, $p^{|V|}q^{B_1}$ is still valid, 
and the EE for the ground states $\ket{\mbox{GS}\,s}$ includes the global constant term $-\gamma=-\log_2 q$. 
All the excitations are deconfined and given by the Wilson line operators, which constitute anyon-like excitations in the interplay with the magnetic fluxes in the ground states.  

\end{itemize}

\subsection{Outlook}
We present some directions for further study:

\begin{itemize}
\item
The toric code model can be extended to a lattice discretizing surfaces with boundary~\cite{freedman,bravyi}. 
It is worth considering a similar extension in the models on graphs presented here.   
In general, it seems nontrivial to divide a graph into bulk and boundary parts. Tree graphs and finite regular lattices are examples in which such a division is possible. 
In the tree graph, vertices with valency 1 are identified as boundaries. In the finite square lattice, vertices with valency 2 or 3 and edges connecting them form the boundary. 
It is interesting to find some other class of graphs such that the division is possible. 
If the models are defined on these graphs with appropriate modifications to the Hamiltonian at the boundary, 
we expect to see the appearance of {\it edge states} 
similar to what happens in the SPT case. This will lead to an extra degeneracy in the number of ground states in addition to the topological and the extensive degeneracy already present. The {\it tree graph} is a special case where we do not expect any topological degeneracy as $B_1=0$ in this case. The interesting thing to note is that these edge states may not result from a fractionalization of a {\it global symmetry} as it happens in the SPT case \cite{nayakSPT}. 
\item
Generalizations of these models to finite non-abelian groups are possible along the lines presented in \cite{pp1}. These are however much harder to analyze.
It is also natural to see if these models can be generalized to other algebras much like the quantum double models of Kitaev. With the machinery developed in \cite{pp1} this might be possible as well. 
\item
Locally the vertex and plaquette operators of the quantum double models satisfy the relations of a {\it quasitriangular Hopf algebra} \cite{majid}. In fact Drinfeld's quantum double construction is tailored to construct such algebras that have the $R$-matrix satisfying the Yang-Baxters equation encoded in them. This gives rise to the anyon excitations which are the IRR's of this algebra \cite{bais}. It would be very interesting to study the way in which this algebra is modified for the operators presented here. Naturally we may expect them to have some generator that can realize the relations of the graph braid groups.
\item
The result obtained here is valid for the same model defined on lattices in arbitrary dimensions, because graphs includes lattices in any dimensions. 
It is expected that these results apply to a broad class of lattice models. 
\item
We have seen that the GSD of these models are a mixture of a topological invariant (first Betti number) and an extensive quantity depending on the lattice size. Moreover apart from the anyonic excitations that account for the topological EE, there are other {\it immobile} excitations that cannot be moved around the graph. These features are very similar to those present in the {\it fracton phases of matter} as discussed in \cite{fr1, fr2, fr3}. We hope to explore more of this connection. 
\item
These systems can also be viewed as stabiliser codes and it is natural to look for an application in quantum computation. However for the case when $q\neq 1$ there are weight 1 symmetries implying that single qubit operations move us within the logical space making them undetectable.
Thus direct application of our models to quantum error correction seems to be restrictive to the case $p=1$. For $p\neq 1$, the $p^{|V|}$ ground states $\ket{\mbox{GS}[\alpha,\beta]}$ with fixed $\beta_L$'s are flipped to one another by 
local perturbations as seen in (\ref{zv_GSab}). From (\ref{ZCL_GSab}), the $p=1$ case seems interesting especially when the lengths of all the closed paths $C_L$ grow as the size of the graph $G$, $|V|$ or $|E|$, increases. 
Then, the ground states $\ket{\mbox{GS}[-,\beta]}$ 
(there is no $\alpha$ parameter when $p=1$) are expected to be robust against local perturbations and useful for error correcting purposes.    
We present such an example in Fig.~\ref{fig:net}. The size of the graph, $|E|$ or $|V|$, grows as $N$ increases. 
Then, both the size of the holes and the number of holes also scale. 
The logarithm of the GSD is subextensive: $\log_2 (\mbox{GSD})=O(N^2)=O\left(|V|^{2/3}\right)=\left(|E|^{2/3}\right)$.
From the quantum computation perspective, $\ket{\mbox{GS}[-,\beta]}$'s are logical qudits. 
Starting from any state of $\ket{\mbox{GS}[-,\beta]}$'s, we can exhaust all the logical qudits by successively acting $Z(C_L)$'s as seen in (\ref{ZCL_GSab}). 
The logical operations grow with the system size just as in the case of surface codes.
The example seems to realize robustness under local perturbations and the growth of the number of logical qudits as $N$ increases.  
%
\begin{figure}[h]
\centering
\captionsetup{width=.8\linewidth}
\begin{tikzpicture}[scale=0.6]
\foreach \x in {-8,-4,0,4,8}
   \draw[-,thick] (\x,-8)--(\x,8);
\foreach \x in {-8,-4,0,4,8}
   \draw[-,thick] (-8,\x)--(8,\x);    
\foreach \x in {-8,-7,-6,-5,-4,-3,-2,-1,0,1,2,3,4,5,6,7,8}
  \foreach \y in {-8,-4,0,4,8} {
   \fill (\x,\y) circle [radius=2pt];
   \fill (\y,\x) circle [radius=2pt];
      }
\draw [<-,thin] (-8,8.7)--(-6.5,8.7);
\draw [->,thin] (-5.5,8.7)--(-4,8.7);
\node(N) at (-6,8.7) {$N$};
\draw [<-,thin] (-8.7,8)--(-8.7,6.5);
\draw [->,thin] (-8.7,5.5)--(-8.7,4);
\node(N) at (-8.7,6) {$N$};
\end{tikzpicture}
\caption{A graph whose shape is a coarse net. The orientations of the edges are suppressed. The size of each hole of the net is $N\times N$, and the number of the holes is $B_1=N^2$. 
$|E|=2N^3+2N^2$ and $|V|=2N^3+N^2+1$. The case $N=4$ is depicted. }
\label{fig:net}      
\end{figure}
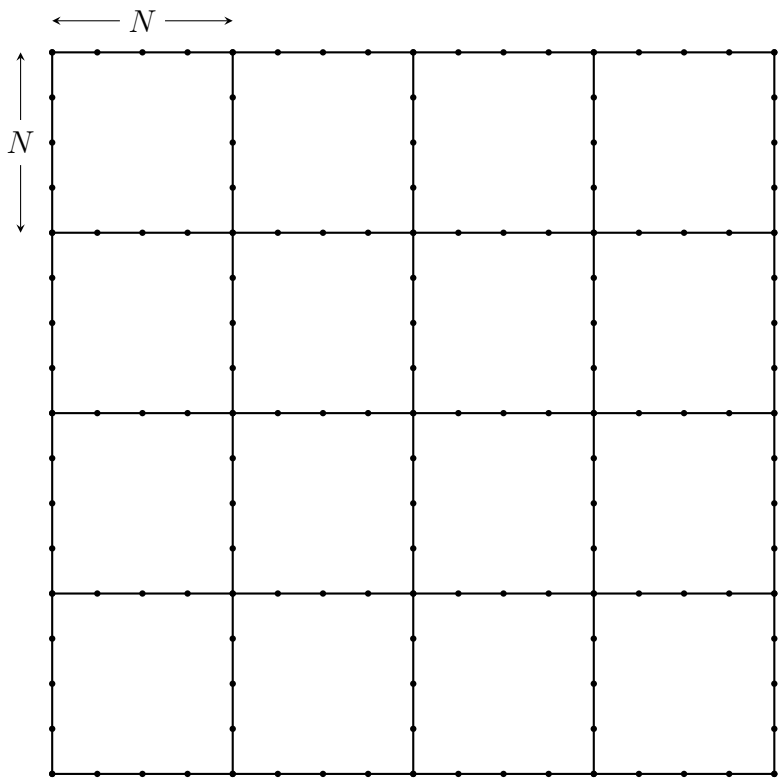
%
\item As a final remark we note that it is possible to go beyond the chain complex/higher gauge formulation presented in \cite{pt1, pt2} with the following example :
\be
A_v = \frac{1}{m}\sum\limits_{j=0}^{m-1}~x_v^{qj}\bX_{L_v}^{qj}, \qquad
B_e = \frac{1}{k}\sum\limits_{j=0}^{k-1} z_{v_1}^{pj}Z_e^{qj}z_{v_2}^{-pj}.
\ee
This model can be thought of as the one where the 0-gauge fields on the vertices act on the 1-gauge fields on the edges, 
\be
\tilde{\partial} : \mathbb{Z}_m \rightarrow \mathbb{Z}_n;~~~ \tilde{\partial}(j) = qj,
\ee
which is a valid homomorphism as $\tilde{\partial}(m) = qm = pn \equiv 0~\textrm{mod}~n$. 
It is opposite to the direction of the homomorphism $\partial$.  
This model has a ground state degeneracy given by
\be
\mbox{GSD} = q^{|E|}\eta,
\ee
with $\eta=\gcd(q,m)$. We hope to analyze these directions in the future.

\end{itemize}

\subsection*{Acknowledgements}
 FS is supported by the Institute for Basic Science in Korea (IBS-R018-D1). 

\appendix
\section{Analogy to quantum field theory with $U(1)$ gauge field and a matter field}
\label{app:analogy}

To gain an intuitive understanding of our model, we consider an analogy to quantum field theory with $U(1)$ gauge field $\vec{A}$ and a complex scalar field $\phi$ for  charged matter. 
$\vec{A}$ means space components of the gauge field, and the time component $A_0$ is supposed to be fixed at $A_0=0$. 

The second term in the Hamiltonian (\ref{H}), $-\sum_{e\in E}B_e$, can be regarded as the kinetic term of the matter field $|\vec{D}\phi|^2$, 
where $\vec{D}$ denotes the covariant derivative associated to the gauge field. 
Then, we can interpret $Z_e$ and $z_v$ as (the exponential of) the gauge field $\vec{A}$ and the matter field $\phi$, respectively. 
(\ref{ZC}) corresponds to the Wilson loop, and $\ket{h_v}\in \cH_v$ ($\ket{i_e}\in \cH_e$) correspond to the eigenstate of the matter field $\phi$ (the gauge field $\vec{A}$). 

As mentioned in section~\ref{sec:vo}, the first term in (\ref{H}), $-\sum_{v\in V}A_v$, imposes the gauge invariance energetically. 
Namely, the gauge symmetry is preserved on the eigenstates of $A_v$ with the eigenvalue 1 for all $v\in V$. 
$A_v$ is analogous to an operator imposing the Gauss law constraint ${\rm div}\,\vec{E}-\rho_{\rm matter}=0$ 
($\vec{E}$ and $\rho_{\rm matter}$ stand for the electric field and the matter charge density, respectively).   
We can interpret $X_e$ and $x_v$ as the electric field (the conjugate momentum of the gauge field) and the conjugate momentum $\pi$ of the matter field. 

The Hamiltonian (\ref{H}) does not possess the plaquette terms of $Z_e$ which corresponds to the gauge kinetic terms 
$\vec{E}^2+ \vec{B}^2$ with $\vec{B}\equiv {\rm rot}\,\vec{A}$ in the gauge field theory.  

Note that the Wilson loop $\exp\left(\mathrm{i}\int_Cd\vec{x}\cdot\vec{A}\right)$ along a loop $C$ in the space creates the unit electric flux along the loop, 
which is seen from its canonical commutation relation with $\vec{E}$: $[A_i(t,\,\vec{x}),\,E_j(t,\vec{x'})]=\mathrm{i}\delta_{ij}\delta(\vec{x}-\vec{x'})$.    

\section{$U(1)$ gauge theory on a circle}
\label{app:U(1)}
In this appendix we briefly review $U(1)$ gauge theory defined on a circle and the property of its vacuum, 
in order to help understanding the ground states in sections~\ref{sec:GSs} and \ref{sec:GSab}. 

For a mathematically well-defined treatment, we also impose the periodic boundary conditions in the time direction $x^0=t\in [0,T]$ as well as the space direction $x^1=x\in [0,L]$. 
After final results are obtained, we can send $T$ to infinity. 
In gauge theory it is sufficient that gauge fields are periodic modulo gauge transformations. 
In general, gauge fields $A_\mu(t,\,x)$ ($\mu=0,1$) satisfy
\be
A_\mu(T,\,x) = A_{\mu}(0,\,x) + \mathrm{i}h(x)\partial_\mu h(x)^{-1}, \qquad 
A_\mu(t,\,L) = A_{\mu}(t,\,0) + \mathrm{i}g(t)\partial_\mu g(t)^{-1},
\ee
where $h(x),\,g(t)\in U(1)$ are transition functions at $t=T$ and $x=L$, respectively. 
In order to obtain topologically nontrivial configurations (nontrivial $U(1)$ bundles over the 2-torus), we take 
\be
h(x)=\exp\left(\frac{2\pi\mathrm{i}}{L}n_xx\right), \qquad g(t)= \exp\left(\frac{2\pi \mathrm{i}}{T}n_tt\right) \qquad (n_x,\,n_t\in \mathbb{Z})
\ee
as an example.
Those with nontrivial $n_x$ and $n_t$ cannot be obtained by continuous deformations from the identity: $h(x)=1$ and $g(t)=1$. 
As discussed in~\cite{tHooft}, we can undo one of the twists, say $g(t)$, 
using a gauge transformation $\Omega(t,\,x)$ such that $\Omega(t,\,0)=1$ and $\Omega(t,\,L)=g(t)$.   
For example, we take
\be
\Omega(t,\,x)=\exp\left(\frac{2\pi\mathrm{i}}{TL}n_ttx\right),
\ee
and then obtain 
\be
A_\mu(T,\,x)=A_\mu(0,\,x)+\delta_{\mu,1}\frac{2\pi\nu}{L}, \qquad
A_\mu(t,\,L)=A_\mu(t,\,0)
\label{AmuBC}
\ee
with $\nu=n_x-n_t$. 

In the boundary conditions (\ref{AmuBC}), we can take the $A_0=0$ gauge. 
Then, topologically nontrivial gauge transformations labelled by an integer $\nu\in \mathbb{Z}$ are given by 
$h_\nu(x)= \exp\left(\frac{2\pi\mathrm{i}}{L}\nu x\right)$ times topologically trivial gauge transformations. 
The topologically trivial gauge transformations are connected to the identity by continuous deformations.  
The configuration space of the gauge field $A_1$ is also divided into the sectors. Namely, configurations satisfying 
\be
A_1(T,\,x)=A_1(0,\,x) + \frac{2\pi\nu}{L}
\label{A1BC}
\ee 
belong to the sector $\nu$. 
Correspondingly the Hilbert space is classified by the topological number $\nu$. 
Given an initial state (at $t=0$) with the topological number $\nu_0$, time evolution of the system under the condition (\ref{A1BC}) 
leads to a final state (at $t=T$) with the topological number $\nu+\nu_0$. 
The vacuum with the topological number $\nu$, $\ket{\Omega_\nu}$ ($\nu\in \mathbb{Z}$), is changed to the one with different $\nu$ by topologically nontrivial gauge transformations. 
However, the $\theta$-vacuum defined by
\be
\ket{\theta}\equiv \sum_{\nu\in\mathbb{Z}}e^{\mathrm{i}\nu\theta}\ket{\Omega_\nu}
\label{thetavac}
\ee
becomes an eigenstate for any gauge transformation. 

From (\ref{A1BC}) it can be seen that $\nu$ is equal to the first Chern number:
\be
c_1\equiv\frac{1}{2\pi}\int_0^Tdt\int_0^Ldx\,F_{01}= \frac{1}{2\pi}\int_0^Tdt\int_0^Ldx\,\partial_0A_1= \frac{1}{2\pi}\int_0^Ldx\,\left[A_1(T,\,x)-A_1(0,\,x)\right]=\nu.
\label{chern}
\ee
This formula indicates that the vacuum in the nontrivial topological sector $\ket{\Omega_\nu}$ has a nontrivial background field strength $F_{01}$. 
Suppose we can take a simply connected domain surrounded by the circle $[0,L]$. 
Magnetic flux penetrating the domain can be expressed as $\Phi(t)=\int^L_0dx\,A_1(t,\,x)$.  Then (\ref{chern}) immediately gives
\be
\Phi(T)-\Phi(0)=2\pi\nu,
\label{Phi}
\ee
which means that the twist (\ref{A1BC}) provides the magnetic flux $2\pi \nu$.  

For a system on a graph as we are discussing in the text, we can consider a subsystem on each closed path $C_L$ analogously to the $U(1)$ theory on a circle here, 
at least regarding the topological structure of the ground states. 

\section{Short review of graph homology}
\label{app:betti}

Algebraic topology helps distinguish topological spaces systematically. The fundamental group and higher homotopy groups classify topological spaces by characterizing the {\it holes} of different dimensions in these spaces but they quickly become hard to interpret as we increase the dimension of the topological space. A commutative alternative to homotopy is given by {\it homology} theory which we are concerned with.

If $X$ is a topological space we can construct a sequence of groups, $H_n(X)$ for $n=0,1,2,\cdots$, termed as the homology groups. 
These are commutative and their rank measures the number of $n$ dimensional holes in $X$. We will illustrate these groups with the simplest example of $X$, a graph. 
Consider the graph shown in Fig. \ref{bettigraph}. 
\begin{figure}[H]
\centering
\captionsetup{width=0.8\linewidth}
\begin{tikzpicture}
%
\fill (0,0) circle [radius=2pt];
\fill (2,0) circle [radius=2pt];
\fill (0,2) circle [radius=2pt];
\draw [->, thick] (0,0)--(1, 0);
\draw [-, thick] (1,0)--(2, 0);
\draw [->, thick] (0,2)--(0, 1);
\draw [-, thick] (0,1)--(0, 0);
\draw [->, thick] (2,0)--(1, 1);
\draw [-, thick] (1,1)--(0, 2);
\draw [->,thick] (2,0) arc (0:45:2.0cm);
\draw [-,thick] (1.42,1.4) arc (45:90:2.0cm);
\node (x) at (-0.3,2) {$x$};
\node (y) at (0,-0.4) {$y$};
\node (z) at (2,-0.4) {$z$};
\node (a) at (-0.3,1) {$a$};
\node (b) at (1,-0.3) {$b$};
\node (c) at (0.8,0.7) {$c$};
\node (d) at (1.75,1.5) {$d$};
\end{tikzpicture}
\caption{A directed graph, $X$ with three vertices, $\{x,y,z\}$ and four edges, $\{a, b, c, d\}$.
}
\label{bettigraph}
\end{figure}
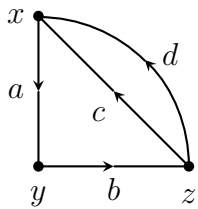
This graph is made up of three vertices $x,~y,~z$ and four edges $a,~b,~c,~d$. The edges are directed as shown in Fig. \ref{bettigraph}. The vertices are also called {\it 0-simplices} and the edges, {\it 1-simplices}. Together the graph $X$ is a {\it simplicial complex}. 
Naturally higher dimensional surfaces correspond to higher simplices but here we restrict ourselves to 0- and 1-dimensional simplices as we are interested in graphs. 

The set of vertices is denoted by $C_0$ and is the free abelian group generated by the vertices $x,y,z$. 
A general element of $C_0$ is $\alpha x + \beta y + \gamma z $ with the coefficients $\alpha, \beta, \gamma$ being numbers in some field which we take to be the integers, $\mathbb{Z}$. 
Likewise, the set of edges is denoted by $C_1$ and is the free abelian group generated by $a,b,c,d$. In the literature the elements of $C_0$ and $C_1$ are called zero- and one-dimensional chains, respectively. 

We now consider a group homomorphism, 
\be
C_1 \overset{\partial_1}{\longrightarrow} C_0,
\nonumber
\ee
which is called the {\it boundary map}. As the name implies it maps the edge in $C_1$ to its boundary in $C_0$. For the case of the graph $X$ in Fig. \ref{bettigraph} we obtain 
%
\be
\partial_1(a) =y - x,\qquad  \partial_1(b)  =  z - y, \qquad
\partial_1(c) = x - z, \qquad  \partial_1(d)  =  x - z.
\ee
%
Clearly the 0-chains $y-x$, $z-y$, etc are the {\it boundaries} of the 1-chains or edges. We can now think of special 1-chains called {\it cycles} whose boundary is null. 
For the graph $X$ in Fig. \ref{bettigraph} we obtain three cycles, $a+b+c$, $a+b+d$ and $c-d$, each of whose boundaries evaluate to 0. A crucial property of the boundary map,
\be
\partial^2=0,
\ee
can be verified by evaluating $\partial_1^2$ in $X$. 

Consider the {\it short exact sequence} 
\be
0 \overset{\partial_2}{\longrightarrow} C_1 \overset{\partial_1}{\longrightarrow} C_0 \overset{\partial_0}{\longrightarrow} 0. 
\ee
The homology groups $H_n$ are defined as
\be
H_n(X) = Z_n/B_n,
\ee
where $Z_n$ are the group of cycles and $B_n$ are the group of boundaries. More precisely $Z_n={\rm Ker}(\partial_n)$ and $B_n={\rm Im}(\partial_{n+1})$. The quotient $Z_n/B_n$ collects $n$-chain cycles that are not boundaries of $n+1$-chains. Thus it is the group generated by the independent $n$-dimensional cycles or holes. 

Thus for the graph $X$ in Fig. \ref{bettigraph} we can compute $H_1 = {\rm Ker}(\partial_1)/{\rm Im}(\partial_2)$. ${\rm Ker}(\partial_1) = \mathbb{Z}\oplus\mathbb{Z}$ is generated by $a+b+c$ and $a+b+d$ 
(two of the obtained three cycles are linearly independent), 
and ${\rm Im}(\partial_2)=0$ as there are no 2-chains for the graph $X$. 
Thus $H_1(X)=\mathbb{Z}\oplus\mathbb{Z}$ essentially enumerates the number of independent one-dimensional cycles in $X$.
The rank of $H_1(X)$ is known as the {\it first Betti number}, $B_1(X)$, which is equal to 2 for the graph $X$.

For a general graph with $|E|$ edges and $|V|$ vertices $H_1 = \left(\oplus \mathbb{Z}\right)^{|E|-|V|+1}$
and hence $B_1=|E|-|V|+1$. From a familiar result in graph theory we identify $|E|-|V|+1$ to be the number of independent cycles of the graph under consideration.

We can also compute $H_0(X)= {\rm Ker}(\partial_0)/{\rm Im}(\partial_1)$ for the graph $X$ in Fig. \ref{bettigraph}. 
Now ${\rm Ker}(\partial_0)=\mathbb{Z}\oplus\mathbb{Z}\oplus\mathbb{Z}$ is generated by $x,y,z$ 
and ${\rm Im}(\partial_1)=\mathbb{Z}\oplus\mathbb{Z}\oplus\mathbb{Z}$ is generated by $y-x, z-y, x-z$. 
To take the quotient we equate each element in $Im(\partial_1)$ to 0 which implies $H_0(X)=\mathbb{Z}$. From this example we can convince ourselves that all vertices in a connected component of a general graph will 
be identified. Thus $H_0$ for a general graph measures the number of connected components of the graph and denotes the {\it zeroth Betti number}. Clearly for the graph $X$ in Fig. \ref{bettigraph}, $B_0(X)=1$.


\end{document}